\documentclass[aps,showpacs,preprintnumbers,amsmath,amssymb]{revtex4}
\usepackage{subeqnarray}
\usepackage{graphicx}
\usepackage{color}
\definecolor{navyblue}{rgb}{0.3,0.3,1}
\definecolor{purple}{rgb}{0.6,0,0.5}
\usepackage[colorlinks=true, pdfstartview=FitV, linkcolor=purple, citecolor=
blue,urlcolor=navyblue]{hyperref}

\begin{document}

\title{Final state interaction in $D^+\to K^-\pi^+\pi^+$ with $K\pi$ I=1/2 and 3/2 channels}
 \date{\today}

\author{K. S. F. F. Guimar\~aes$^a$, O. Louren\c{c}o$^b$, W. de Paula$^c$,  T. Frederico$^c$,  A. C. dos Reis$^d$}

\affiliation{$^a$Instituto de Astronomia, Geof\'isica e Ci\^encias Atmosf\'ericas, 
05508-900, S\~ao Paulo, SP, Brazil  \\
$^b$Departamento de Ci\^encias da Natureza, Matem\'atica e Educa\c
c\~ao, CCA, Universidade Federal de S\~ao Carlos, 13600-970, Araras, SP, Brazil\\
\mbox{$^c$ Instituto Tecnol\'ogico de
Aeron\'autica, 12228-900, S\~ao Jos\'e dos Campos, SP, Brazil }\\
  \mbox{$^d $Centro Brasileiro de Pesquisas F\'isicas,
22290-180, Rio de Janeiro, RJ, Brazil}}

\begin{abstract}
The final state interaction contribution  to $D^+$ decays is computed for the
$K^-\pi^+\pi^+$ channel within a light-front relativistic three-body model for the final
state interaction. The rescattering process between the kaon and two pions in the decay
channel is considered. The off-shell decay amplitude is a solution of a four-dimensional
Bethe-Salpeter equation, which is decomposed in a Faddeev form. The projection onto the
light-front of the coupled set of integral equations is performed via a quasi-potential
approach. The S-wave $K\pi$ interaction is introduced in the resonant isospin $1/2$ and
the non-resonant isospin $3/2$ channels. The numerical solution of the light-front
tridimensional inhomogeneous integral equations for the Faddeev components of the decay
amplitude is performed perturbatively. The loop-expansion converges fast, and the
three-loop contribution can be neglected in respect to the two-loop results for the
practical application. The dependence on the model parameters in respect to the input
amplitude at the partonic level is exploited and the phase found in the experimental
analysis, is fitted with an appropriate choice of the real weights of the isospin
components of the partonic amplitude. The data suggests a small mixture of total isospin
$5/2$ to the dominant $3/2$ one. The modulus of the unsymmetrized decay amplitude, which
presents a deep valley and a following increase for $K\pi$ masses above $1.5$~GeV, is
fairly reproduced. This suggests the assignment of  the quantum numbers  $0^+$ to the
 isospin 1/2 $K^*(1630)$ resonance.
\end{abstract}

\pacs{ 13.25.Ft,11.80.Jy,13.75.Lb}

\maketitle
\section{Introduction}

Weak decays of heavy flavoured hadrons  provide unique opportunities to
probe the interplay of the electroweak theory and Quantum Chromodynamics (QCD).
The weak part of these decays involve short-distance transitions at the quark-level,
whereas the hadron formation is governed by the long-distance, low-energy strong
interactions.

Due to the non-perturbative character of the strong interactions involved in
heavy flavour decays, the hadronization is not calculable from first principles.
In the kaon sector, chiral perturbation methods are applicable, given the small
value of the $s$ quark mass. In the opposite extreme, the mass of the $b$ quark
is heavy enough to allow for reliable calculations based on effective field theories.
The charm quark is in between these two cases, which makes the computation of
decay rates a challenging task.

The study of the {\em C}harge-{\em P}arity ({\em CP}) violation \cite{CPV1, CPV2}
is an important example where the hadronic part of the decay amplitude needs to be
quantitatively understood. {\em CP}) violation is phenomenon where manifestations
of new physics are expected. In the Standard Model (SM), {\em CP} violation processes
are related to the  complex phase in \textit{Cabibbo-Kobayashi-Maskawa} matrix
(CKM)~\cite{CKM1,CKM2}, which describes the mixture between different generations of
quarks. SM predicts very small {\em CP} violation effects in charm decays, in spite
of large uncertainties. This makes charm decays a very interesting place to search
for new sources of {\em CP} violation.
New physics would introduce additional {\em CP}-violating phases, but disentangling
these from the SM {\em CP} violation require the control of the overwhelming strong
phases.

We emphasize the advantages of the experimental investigation of the three-body
charm meson decays. These decays are, in general, dominated by resonant intermediate
states, with a small non-resonant component \cite{Bianco}. With three-body decays
one can search for local {\em CP} violation effects, but the description of the
decay dynamics  requires the understanding of hadronic effects such as the
three-body final state interactions and the role of the S-wave component.

In this paper we address the issue of three-body final state interactions  (FSI) in the
decay $D^{+}\rightarrow K^{-}\pi^{+}\pi^{+}$\footnote{Charge conjugation is implicit
throughout this paper.}, with emphasis on the S-wave component
of the $K^{-}\pi^{+}$ amplitude. This channel is chosen for several reasons: it is
abundant, being studied by different experiments like E791 \cite{Aitala12,Aitala3},
FOCUS \cite{FOCUS1,FOCUS2} and CLEO \cite{CLEO};
it has a dominant S-wave component and a small non-resonant amplitude; it allows
the continuously study of the $K\pi$ S-wave amplitude from threshold, at $633$~MeV/c$^2$,
up to $1.7$~GeV/c$^2$, covering the whole elastic regime. With the
$D^{+}\rightarrow K^{-}\pi^{+}\pi^{+}$ decay one can fill the gap of the existing data
on $K\pi$ scattering from the LASS experiment \cite{LASS}
(LASS data for the $K\pi$ scattering starts only at 825~MeV/c$^2$).

The resonant structure of  three-body decays are determined by the analysis of the
Dalitz plot \cite{Dalitz}. In this two-dimensional diagram,
the probability density of a pseudo-scalar particle $P$,
decaying into three pseudoscalar particles ($d_{1},d_{2},d_{3}$), is given by
\begin{eqnarray}
 d\Gamma(P \to d_{1}d_{2}d_{3})\propto
 \frac{1}{M^{3}_{P}}\, |\mathcal{M}(s_{12},s_{13})|^{2} \,ds_{12}\, ds_{13}
\end{eqnarray}
where $M_P$ is the mass of the parent particle. The phase-space density,
${M^{-3}_{P}}$, is constant, so the structures reveal the decay dynamics, forming the
resonances, which are also affected by final state interactions. The goal of the
Dalitz plot analysis is to determine the matrix element $\mathcal{M}(s_{12},s_{13})$.

The Dalitz plot analysis of the $D^{+}\rightarrow K^{-}\pi^{+}\pi^{+}$ was performed by
different experiments, such as MARK III \cite{Adler1987a,Adler1987b,Adler1987c,Adler1987d},
NA14 \cite{Alvarez,Alvarez2}, E691 \cite{Anjos,Anjos2}, E687 \cite{Frabetti,Frabetti2}, E791\cite{Aitala12}
and FOCUS \cite{FOCUS1,FOCUS2}, using different decay models. These decay models differ
in the way the S-wave is described: the sum of Breit-Wigners plus a constant nonresonant
term, refered to as the Isobar Model, the K-matrix formalism and a model independent
partial wave analysis (MIPWA), to which we give special attention.

The MIPWA technique, developed by E791 \cite{Aitala2006}, is
intended to extract, in a independent way, the S-wave
$K\pi$ amplitude of the $D^{+}\rightarrow K^{-}\pi^{+}\pi^{+}$ decay. In the MIPWA, the
S-wave $K\pi$ amplitude is a generic function, $A_{0}(s) = a_{0}e^{i\phi_{0}(s)}$, given
by the fit of the Dalitz plot. The P and D wave are determined according the Isobar Model.
Although the MIPWA is the most model-independent approach, the extraction of the phase is
an inclusive measurement, comprising different isospin amplitudes and FSI.

As a matter of fact, the comparison between the S-wave from
scattering and from $D$ decays show important differences which need to be understood.
In addition to an overall shift of approximately 150 degrees, the two amplitudes
have different shapes.

The S-wave $K\pi$ amplitude depends on the isospin and orbital angular momentum of
the system. There are two isospin states possible for this system, namely, $I=1/2$ and
$I=3/2$. In the case of the LASS experiment, it was shown that resonances and the
corresponding scattering amplitude poles are present only in the isospin 1/2 channel, as
verified in the analysis of the phase $\delta_{I=1/2}(m_{K\pi})$ \cite{Alberto}. It is
expected that this phase would be common to all processes having  a $K\pi$ system,
in the absence of rescattering involving other particles in the final state.
This should be valid to all angular momentum states,  according to
the Watson theorem \cite{Watson}.

The S-wave phase-shift obtained from the  $D^+\to K^-\pi^+\pi^+$ decay with the MIPWA
(FOCUS and E791) differ from that obtained from $K\pi$ scattering (LASS). There is an
energy dependent discrepancy that cannot be cured by any combination of $\delta_{I=1/2}$
and $\delta_{I=3/2}$. Indeed, up to an overall shift of $\sim150^{\circ}$ ,
such an energy dependence was reproduced quite nicely
below $K^*_0(1430)$ in a chiral three-body model of the $K\pi\pi$ decay with S-wave
$K\pi$ interaction, in the resonant isospin $1/2$ channel and computed up to
two-loops~\cite{MagPRD11}.  We should mention that a previous attempt \cite{BoitoPRD09,Boito:2010qe}
to describe the decay $D^{+}\rightarrow K^{-}\pi^{+}\pi^{+}$ considering only two-body FSI
(no 3-body FSIs and factorization of the weak vertex) was also quite successful
phenomenologically below $K^*_0(1430)$.

Our aim is to further explore  theoretically the three-body final state interaction in the
$D^{+}\rightarrow K^{-}\pi^{+}\pi^{+}$ decay. The motivation of our study is the
possibility of  three-body rescattering in $D^{+}\rightarrow K^{-}\pi^{+}\pi^{+}$ decay
for $K\pi$ interactions in both isospin channels, while fitting the LASS data in the whole
kinematical region of the experiment up to $1.89$~GeV. Our study is based in  a
relativistic model for the three-body final state interaction in $D^{+}\rightarrow
K^{-}\pi^{+}\pi^{+}$ decay, starting with the three-meson Bethe-Salpeter equation
\cite{karinnpb,pos,MagPRD11}.

In the model developed here, the decay amplitude is separated into a smooth term and a
three-body fully interacting contribution, which is factorized in the standard two-meson
resonant amplitude times a reduced complex amplitude for the bachelor meson, that carries
the effect of the three-body rescattering mechanism. The off-shell bachelor reduced
amplitude is a solution of an inhomogeneous Faddeev type integral equation, that has as
input the S-wave isospin $1/2$  and $3/2$ $K^{-}\pi^{+}$ transition matrix. The
theoretical contribution of the present work is to use in the three-body rescattering
equations the S-wave two-body $K\pi$ amplitude in both isospin states, $1/2$ and $3/2$,
fitted up to $1.89$~GeV. We neglect the  interaction between the identical charged pions.

The three-body model of the decay amplitude is recasted in a Bethe-Salpeter like
equation, which is conveniently rewritten in terms of a Faddeev expansion. The
contribution of the final state interaction in the three-body decay of a heavy-meson in
our model of the S-wave $K\pi$ transition amplitude is encoded by a bachelor amplitude
associated with each Faddeev component of the full decay amplitude. The bachelor function
modulates the $K\pi$ scattering amplitude in the final decay channel and in general
carries a phase. The advantage of using the Faddeev decomposition of the decay amplitude,
is that (i) the integral equation for the bachelor function has a connected kernel, and
(ii) the kernel is written in terms of the two-body scattering amplitude directly, instead
of the potential. We use a parametrization of  the $K\pi$ scattering amplitude in $I=1/2$
and $3/2$, which is input to the bachelor integral equations, and constitutes one source
of the energy dependence seen in the $D^{+}\rightarrow K^{-}\pi^{+}\pi^{+}$ S-wave phase
shift, besides the phase of the $K\pi$ amplitude. Technically, we perform the light-front
projection of the equations
\cite{sales00,MarPRD07,MarPoS08,MarFBS08,MarPRD08,FredFBS11,FreFBS14}, to simplify the
numerical computation of the observables by three-dimensional integrations. These
techniques are well exemplified in the reviews of applications of light-front field theory
to nuclear and hadron physics \cite{Karmanov1,brodsky}. In particular, we should mention
the application of light-front quantization to describe
three-body systems, see e.g. \cite{BakNPB79,FudPRC87,FrePLB92,AdhAP94,CarPRC03}.

The work is organized as follows. In Sec.~\ref{sec:kpiampl}, we present our fitting
model for the $K\pi$ $I=1/2$ S-wave phase-shift up to about $1.89$~GeV of the LASS data
\cite{LASS}. In the following sections, the relativistic formalism to compute the
contribution of three-body final state interaction in heavy-meson decays is developed.
In Sec.~\ref{sec:FSIcov}, we present the derivation of a covariant and four-dimensional
Bethe-Salpeter equation for the three-body decay with rescattering effects. In
Sec.~\ref{sec:FSILF}, we present the light-front projection technique and derive the
three-dimensional equations for the bachelor amplitude. In Sec.~\ref{LFMDDecay}, the
isospin projection of the LF equations for the bachelor amplitudes derived in the
preceding section is performed.  The perturbative solution of the LF integral equations
are constructed in Sec.~\ref {pertsol} for the bachelor amplitude up to three-loops,
namely, up to terms in third order in the two-body transition matrix to check convergence.
In Sec.~\ref{sec:results} the numerical results for the $D^{+}\rightarrow
K^{-}\pi^{+}\pi^{+}$ with three-body final state interaction and $K\pi$
interactions in $I=1/2$ and $3/2$ states are presented. In Sec.~\ref{conclusion}, we
summarize the main contributions of this work to both the experimental and theoretical
analysis of the $D^{+}\rightarrow K^{-}\pi^{+}\pi^{+}$ decay.

\section{$K\pi$ S-wave amplitude}
\label{sec:kpiampl}

The S-wave amplitudes of the $K\pi$ elastic scattering in the resonant $I_{K\pi}=1/2$ and
the non-resonant one  $I_{K\pi}=3/2$ states are the inputs of our model of the $3\to 3$
T-matrix, which brings the final state interaction between the three mesons to the
$D^+\to K^-\pi^+\pi^+$ decay. As we already mentioned, the interaction of the
identical pions is neglected. Here, we just follow \cite{karinnpb,FreFBS14} for the
parametrization of the LASS data \cite{LASS} in the S-wave resonant $I_{K\pi}$=1/2
channel. In addition to the $K^*_0(1430)$, we use the resonances $K_0^*(1630)$
(in Particle Data Group \cite{PDG} there is no assignment of spin to $K(1630)$)
and $K_0^*(1950)$). The
lowest resonance and broad one $K_0^*(800)$ comes with the effective range parameters.
In Ref.~\cite{MagPRD11}, it was the result of the low energy chiral dynamics and
unitarity, appearing naturally as a pole in the S-channel.

The motivation to include the higher radial excitations of $K^*_0$ comes from recent
proposal to interpret the scalar meson family ($f_0$) as radial excitations of the
$\sigma$ meson as proposed in Refs.~\cite{dePaulaPLB10,dePaula:2010yu}. This result was
obtained by using a Dynamical AdS/QCD model\cite{dePaula:2008fp}, where the backreaction
between the dilaton field and a deformed anti-de Sitter metric is taken into
account. Using a different approach, in \cite{MasPRD12} it was also proposed a systematics of
radial Regge trajectories for light scalars, which couples these resonances to the $\pi\pi$
channels. By analogy, if these analyses are
extended to the strange sector it would  suggest a mass spectrum ($M^2\times n$) for the
kappa family with a rough slope of $\sim 0.6$~GeV$^{2}$,  and also the decay of these mesons in the $K\pi$ S-wave
$I_{K\pi}=1/2$ channel. The fitting of the LASS data in this isospin channel is the main
reason to use more resonances, namely, $K_0^*(1630)$ and $K_0^*(1950)$  besides $K_0^*(1430)$.
Being conservative, these further resonances can be considered at the moment as a
practical way to fit the data in the whole kinematical range up to $1.89$~GeV.

The parametrization of our relativistic model of the \mbox{S-wave} $I_{K\pi}=1/2$
scattering amplitude extends the one used in Ref.~\cite{BABAR}, where we introduce also
$K_0^*(1630)$ and $K_0^*(1950)$. The relativistic scattering amplitude as a function of
$M^2_{K\pi}$ is written in terms of the S-matrix ($S^{\mbox{\tiny{1/2}}}_{K\pi}$) as:
\begin{eqnarray}
\tau_{\mbox{\tiny{1/2}}}\left(M^2_{K\pi}\right)=4\pi\,\frac{M_{K\pi}}{k}
\left(S^{\mbox{\tiny{1/2}}}_{K\pi}-1\right) \label{tau12}
\end{eqnarray}
where
\begin{eqnarray}
S^{\mbox{\tiny{1/2}}}_{K\pi}= \frac{k\cot\delta+i \,k}{k\cot\delta-i
\,k}\, \prod_{r=1}^3 \frac{M^2_r-M^2_{K\pi}+i z_r \bar\Gamma_r}
 {M^2_r-M^2_{K\pi}-i z_r\Gamma_r}
\label{skpi12}
\end{eqnarray}
and $z_r=k\,M^2_r/(k_r\,M_{K\pi})$, with the  c. m. momentum of each meson of the $K\pi$
pair given by
\begin{eqnarray}
k=\left[\left(\frac{M^2_{K\pi}+m_\pi^2-m_K^2}{2\,M_{K\pi}}
\right)^2-m_\pi^2\right]^{1/2}  \ .\label{kcm}
\end{eqnarray}

For each resonance, we associate the parameters $M_r$, $\Gamma_r$, $\bar \Gamma_r$ and
$k_r$. The momentum $k_r$ corresponds to Eq.~(\ref{kcm}) at the resonance position. The
inelasticity in $K\pi$ \mbox{S-matrix} comes by allowing $\bar\Gamma_r$ and $\Gamma_r$
distinct, such that $-\Gamma_r<\bar\Gamma_r<\Gamma_r$. The resonance parameters
$(M_r,\Gamma_r,\bar\Gamma_r)$ in GeV for $K_0^*(1430)$, $K_0^*(1630)$ and $K_0^*(1950)$
are (1.48,0.25,0.25), (1.67,~0.1,0.1) and (1.9,~0.2,~0.14), respectively
\cite{karinnpb,FreFBS14}. The non-resonant component of the S-matrix is parameterized by
the effective range expansion:
\begin{eqnarray}
k\cot\delta=\frac{1}{a}+\frac12 r_0\, k^2 \label{effexp}
\end{eqnarray}
with $a=1.6$~GeV$^{-1}$ and $r_0=3.32$~GeV$^{-1}$.
\begin{figure}[!htb]
\centering
\includegraphics[scale=0.5]{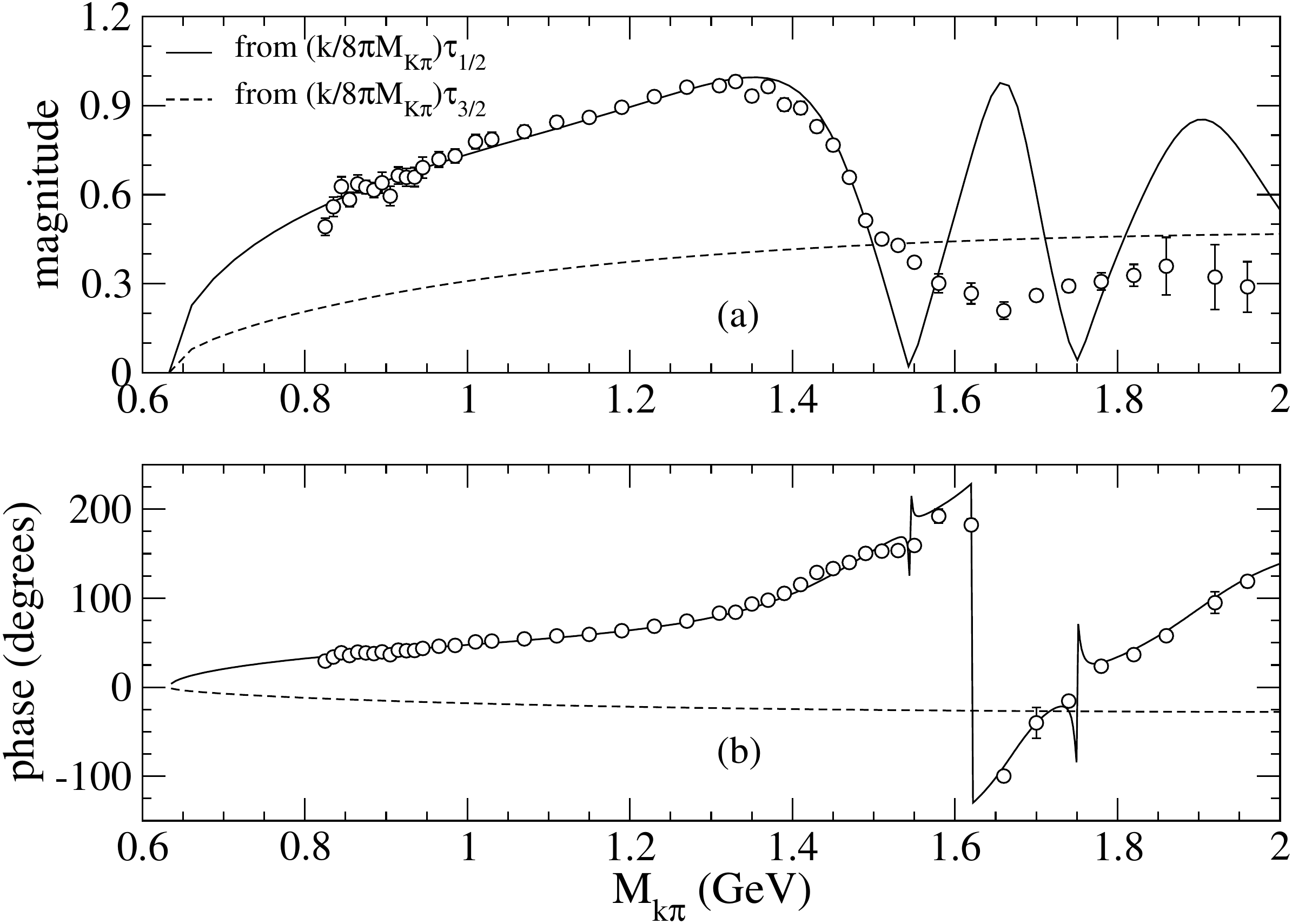}
\caption{Magnitude (a) and phase (b) obtained from both, $I=1/2$, and $I=3/2$ $K\pi$
S-wave amplitude, Eq.~(\ref{tau12}) and Eq.~(\ref{tau32}), respectively. Circles: LASS
phase-shift data \cite{LASS} for $I_{K\pi}=1/2$. }
\label{tau1232}
\end{figure}

The S-wave $I=3/2$ K$\pi$ amplitude is given by
\begin{eqnarray}
\tau_{\mbox{\tiny{3/2}}}\left(M^2_{K\pi}\right)=4\pi\,\frac{M_{K\pi}}{k}
\left(S^{\mbox{\tiny{3/2}}}_{K\pi}-1\right), \label{tau32}
\end{eqnarray}
where
\begin{eqnarray}
S^{\mbox{\tiny{3/2}}}_{K\pi}= \frac{k\cot\delta+i \,k}{k\cot\delta-i
\,k},
\end{eqnarray}
where the effective range expansion of $k\cot\delta$ comes from Eq.~(\ref{effexp}), and
parameters $a=-1.00$~GeV$^{-1}$ and $r_0=-1.76$~GeV$^{-1}$ from Ref. \cite{estabrooks}.
The relative momentum of the $K\pi$ pair is written in Eq.~(\ref{kcm}).

The results from the three-resonance model Eq.~(\ref{skpi12}) are shown in
Fig.~\ref{tau1232} up to $2$~GeV. The $I_{K\pi}=1/2$ S-wave phase-shift is compared to the
LASS. We privileged the fit of the phase-shift and the model parametrization from \cite{karinnpb,FreFBS14} is able to reproduce the
LASS data for the phase reasonably well. The results of the parametrization for  $|S^{\mbox{\tiny{1/2}}}_{K\pi}-1|/2$ as shown in the upper panel of
Fig. ~\ref{tau1232}, which reproduce the data up  to about $K^*(1430)$, present structure not observed in the LASS phase-shift analysis.

\begin{figure}[!htb]
\centering
\includegraphics[scale=0.5]{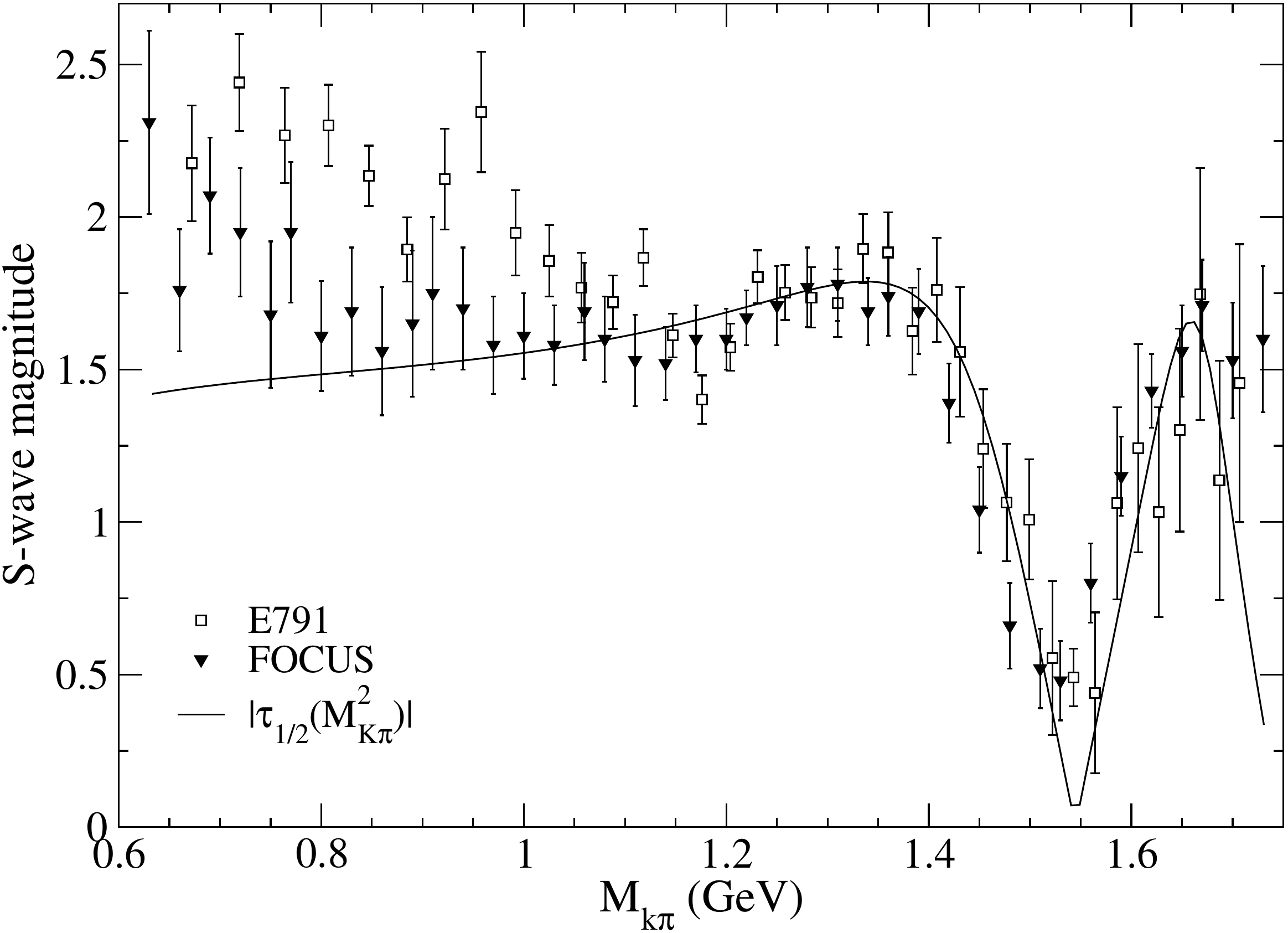}
\caption{Magnitude of the $I_{K\pi}=1/2$ S-wave amplitude as a function of
the $K\pi$ mass. Solid line:
$|\tau_{\mbox{\tiny{1/2}}}\left(M^2_{K\pi}\right)|$ from Eq.~(\ref{tau12}) with arbitrary
normalization.  The data come from
the phase-shift analysis of E791 (empty squares)~\cite{Aitala12,Aitala3} and FOCUS
collaboration (full inverted triangles)~\cite{FOCUS1,FOCUS2}. } \label{tau12mod}
\end{figure}

On the other hand as shown in Fig.~\ref{tau12mod}, the phase-shift analysis for the  $D^+\to K^-\pi^+\pi^+$ decay from  E791~\cite{Aitala12,Aitala3}
and  FOCUS ~\cite{FOCUS1,FOCUS2} collaborations, considering the dominance of this isospin channel in the final state interaction of this
decay \cite{MagPRD11}, suggest that the magnitude from the model parametrization (\ref{tau12}), with the structure shown in Fig. ~\ref{tau1232}  may be possible.
The deep minimum observed in Fig.~\ref{tau12mod}  around 1.53 GeV, is consistent with the zero  of $|\tau_{\mbox{\tiny{1/2}}}\left(M^2_{K\pi}\right)|$, as clearly
depicted in the figure. As we are going to show in detail by calculations of three-body final state interactions in sections \ref{pertsol} and \ref{sec:results}, this feature is kept.

To be complete both isospin $1/2$ and $3/2$ are shown  in Fig. ~\ref{tau1232} for
comparison, and close to the minima of the magnitude of the $I_{K\pi}=1/2$ amplitude, the
$3/2$ one becomes important, just anticipating what would come from the $D$ decay. The
data for $I_{K\pi}=3/2$ is not shown as the effective range parametrization is the fit of
the phase-shifts of this channel already presented in \cite{estabrooks}.

\section{$D^+\to K^-\pi^+\pi^+$ decay with FSI}
\label{sec:FSIcov}

\begin{figure}[!htb]
\centering
\includegraphics[scale=0.6]{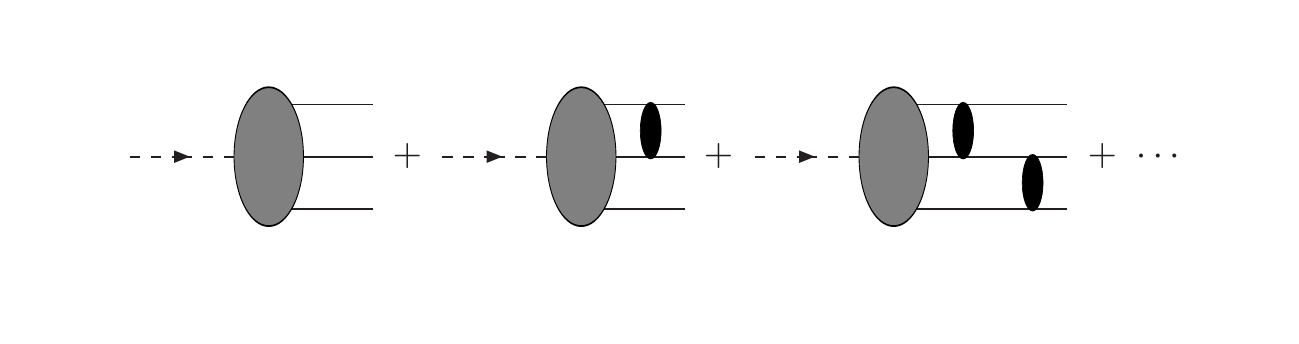}
\vspace{-0.5cm}
\caption{ $D$  decay process into $K\pi\pi$ in the three-body ladder approximation.
The source term (partonic amplitude) is represented by the gray blob. The  fully off-shell $K\pi$ transition matrix is represented by the black blob.}
\label{ladder}
\end{figure}

The collisions between the mesons in the  final state of the $D^+\to K^-\pi^+\pi^+$   is represented
 diagrammatically  in Fig.~\ref{ladder}. The rescattering series is summed up in the $3\to
3$ transition matrix, which composes the
full decay amplitude as (see \cite{MagPRD11}):
\begin{eqnarray}
{\cal A}(k_{\pi},k_{\pi^\prime})=D(k_{\pi},k_{\pi^\prime}
)
+\int \frac{d^4q_\pi d^4q_{\pi^\prime}}{(2\pi )^8}T(k_{\pi},
k_{\pi^\prime};q_{\pi},q_{\pi^\prime})S_\pi(q_\pi)\,S_{\pi}(q_{\pi^\prime})S_K(K-q_{\pi^\prime}-q_{\pi})
D(q_\pi,q_{\pi^\prime}) \ ,\label{D}
\end{eqnarray}
where the momentum of the pions  are
$k_\pi$ and $k_{\pi^\prime}$.

The source of the mesons in the final state is
given by the partonic amplitude expressed by the function
$D(k_\pi,k_{\pi^\prime})$, which is the first term of (\ref{D}) and the gray blob
in Fig. \ref{ladder}. It corresponds to a smooth amplitude given by the direct partonic decay amplitude
determined by short-distance physics.

The second term of  (\ref{D}) brings the long range physics, which is represented by the
sum of rescattering diagrams in the figure, has the $3\to 3$
transition matrix $T(k_{\pi},k_{\pi^\prime};q_{\pi},q_{\pi^\prime})$ convoluted with the source term,
including the off-shell mesonic Feynman propagators  $S_i(q_i)=i(q_i^2-m_i^2+i \varepsilon)^{-1}$ , where the masses are
 $m_i$ $(i=\pi,K,\pi^\prime)$ and the self-energies are disregarded. In the approximation considered in our work, the
 $3\to 3$ transition matrix sums the connected  scattering series from ladder graphs. All possible  $2\to 2$ collision terms  are
summed up in the $K\pi$ transition matrix,  represented by the black blobs in Fig. \ref{ladder}.
As a matter of fact, in the model we develop the T-matrix operator acts on the isospin space of the
$K\pi\pi$ system, while ${D}(k_{\pi},k_{\pi^\prime})$ is an amplitude in the isospin space of the $K\pi\pi$ system.


\subsection{Three-body Bethe-Salpeter approach }
The final state interaction between the mesons in the three-body decay channel, are
given by the full three-body T-matrix. It is a solution of the Bethe-Salpeter (BS)
equation, which will be written in the Faddeev form. We consider spinless particles,
disregard self-energies and three-body irreducible diagrams. Under these assumptions,
the interactions between the mesons are assumed to be only due to two-body interactions.
To be concise the momentum dependences will be omitted in the discussion below.

The three-particle BS equation for the T-matrix can be
written as
\begin{eqnarray}
T=\sum V_{i} + \sum V_{i}G_0\,T , \label{eq1}
\end{eqnarray}
where the sum runs over the three two-body subsystems $i=(j,k)$. Formally, the potential
in the four-dimensional equation is built by multiplying the two-body interaction
$V^{(2)}_{jk}$ from all two-particle  irreducible diagrams in which particles \textit{j
}and \textit{k} interact, and by the inverse of the individual propagator of the spectator
particle $i$, $S_i$
\begin{eqnarray}
V=\sum_{i=1}^3 V_i\ ;~~V_i=V_{(2)jk}S^{-1}_i~.\label{presc}
\end{eqnarray}
 The propagator of particle $i$ is $S_{i}=\imath\left [ k_{i}^{2}-m_i^2+\imath
\epsilon\right]^{-1}$, $k_{i}$ being its four-momentum.
The three-particle free Green's function is
\begin{eqnarray}
G_0=S_{i}S_{j}S_{k}\, . \label{eq2}
\end{eqnarray}
Eq.~(\ref{eq1}) can now be rewritten in the
Faddeev form.  The transition matrix is decomposed as
$T=T^{1}+T^{2}+T^{3}$ with the components $T^i=V_{i}+V_{i}\,G_0\,T$.

The relativistic generalization of the connected Faddeev equations  is
\begin{eqnarray}
T^{i}=T_i+T_{i}G_0\left(T^{j}+T^{k}\right)
, \label{eq4}
\end{eqnarray}
where the two-body T-matrices are solutions of
\begin{eqnarray}
T_i=V_{i}+V_{i}G_0T_{i} , \label{eq5}
\end{eqnarray}
within the three-body system. The full $3 \to 3$ ladder scattering
series is summed up by solving the integral equations for the
Faddeev decomposition of the scattering matrix. Therefore, the three-body
unitarity holds for the 3$\to$3 transition matrix built from the solution
of the set of Faddeev equations (\ref{eq4}) below the threshold of particle production
from two-body collisions, where the two-body amplitude is unitary.

The full decay amplitude, Eq.~(\ref{D}), can be decomposed according to Eq.~(\ref{eq4})
as
\begin{eqnarray}
{\mathcal A}=D+ \sum D^i \ , \label{d1}
\end{eqnarray}
where the Faddeev components of the decay vertex are
\begin{equation}
D^i=T^i\,G_0\,D \ . \label{d2}
\end{equation}
They are solutions of the connected equations
\begin{eqnarray}
D^{i}=D_i+T_{i}G_0\left(D^{j}+D^{k}\right)
, \label{eq40}
\end{eqnarray}
with
\begin{eqnarray}
D_i= T_{i}\,G_0\,D
. \label{eq41}
\end{eqnarray}
The Faddeev equations for the decay vertex, Eqs.~(\ref{eq40})-(\ref{eq41}) are general
once self-energies and three-body irreducible diagrams are disregarded.  In the
following they will be particularized to allow a separable form of the three-body
decay amplitude.

\subsection{s-channel two-meson amplitude}

The matrix elements of the two-particle transition matrix is assumed to
depend only on the Mandelstam \mbox{s-variable} and, within the three-body
system, they read
 \begin{eqnarray}
T_{i}(k'_j,k'_k;k_j,k_k)=(2\pi)^4\tau_i(s_{i}) \,S^{-1}_i(k_i)\,\delta(k'_i-k_i)
\ , \label{eq42}
\end{eqnarray}
where a delta of four-momentum conservation has been factorized out. The S-wave scattering
amplitude $\tau_i(s_{i})$ of particles $i$ and $j$, depends on the Mandelstam variable
$s_{i}=(k_j+k_k)^2$. The three-body unitarity in our formulation is maintained, once the amplitude
$\tau(s)$ is unitary.

Introducing Eq.~(\ref{eq42}) in Eqs.~(\ref{eq40})-(\ref{eq41}), one gets that
\begin{eqnarray}
D^{i}(k_j,k_k)= \tau_i(s_{i})\xi^i(k_i)
, \label{eq43}
\end{eqnarray}
where
\begin{eqnarray}
\xi^i(k_i)=\xi_0^i(k_i)+\int \frac{d^4q_j d^4q_{k}}{(2\pi )^4}\delta(k_i-q_i)
S_j(q_j)S_k(q_k)\left(D^{j}(q_k,q_i)+D^{k}(q_i,q_j)\right) \, ,
\label{eq44}
\end{eqnarray}
and
\begin{equation}
\xi_{0}^i(k_i)=\int \frac{d^4q_j}{(2\pi )^4}S_j(q_j)S_k(K-k_i-q_j)
\,D(q_i,q_j) \, , \label{eq45}
\end{equation}
with $q_k=K-k_i-q_j$. One can simplify the form of Eq.~(\ref{eq44}) by using the
separation of the momentum dependences given by Eq.~(\ref{eq43}),
\begin{eqnarray}
\xi^i(k_i)=\xi_{0}^i(k_i)+\int \frac{d^4q_j d^4q_{k}}{(2\pi )^4}\delta(k_i-q_i)
S_j(q_j)S_k(q_k) \left(\tau_j(s_{j})\xi^j(q_j)+\tau_k(s_{k})\xi^k(q_k)\right)  \ ,
\label{eq46}
\end{eqnarray}
and, integrating the $\delta$'s, the formula is simplified to
\begin{multline}
\xi^i(k_i)=\xi^i_0(k_i)+\int \frac{d^4q_j }{(2\pi )^4}S_j(q_j)S_k(K-k_i-q_k)\tau_j(s_{j})\xi^j(q_j) 
+\int \frac{d^4q_k }{(2\pi )^4} S_j(K-k_i-q_k)S_k(q_k)\tau_k(s_{k})\xi^k(q_k) \ .
\label{eq47}
\end{multline}
The separable form of the two-body T-matrix allows to simplify the integral equation
for the Faddeev components of the vertex function, reducing it to a four-dimensional
integral equation in one momentum variable.

The full decay amplitude considering the final state interaction computed with
Eq.~(\ref{eq43}) reduces to the expression
\begin{eqnarray}
{\mathcal A}_0(k_i,k_j)=D(k_i,k_j)+\sum_\alpha\tau(s_{\alpha})\xi^\alpha(k_\alpha) \ ,
\label{eq48}
\end{eqnarray}
where all the  mesons in the three-body decay channel interact. The subindex in ${\cal
A}_0$ just denotes the s-wave two-meson scattering.

The complex function  $\xi(k_i)$ in Eq.~(\ref{eq48}) carries the three-body rescattering
effect by an amplitude and phase depending on the bachelor meson on-mass-shell momentum,
while $\tau(s_{i})$ takes into account two-meson resonances. In the particular
case of the $D^+\to K^-\pi^+\pi^+$ decay, and assuming that the identical pions
do not interact, Eq.~(\ref{D}) reduces to Eq.~(\ref{eq48}) under the assumption that the
matrix elements of the $K\pi$ transition matrix depend only on the Mandelstam
\mbox{s-variable}.

\subsection{$D^+\to K^-\pi^+\pi^+$ problem}

The $K\pi\pi\to K\pi\pi$ rescattering process is accounted by the $D^\pm$ decay amplitude
expressed by Eq.~(\ref{eq48}), where the bachelor amplitudes $\xi(k)$ are solutions of the
connected Faddeev-like equations~(\ref{eq47}). Furthermore, we simplify the problem and
disregard the interaction between the equal charged pions. The effective S-wave
interaction between the kaon and pion is local on the fields with the $K\pi$ scattering
amplitude $\tau_i(M^2_{K\pi})$ parameterized to reproduce the $K\pi$ S-wave phase-shift
in the isospin $1/2$ and $3/2$ channels from the LASS experiment \cite{LASS}, as presented
in Sec.~\ref{sec:kpiampl}.

The model assumptions for the $D^+\to K^-\pi^+\pi^+$ decay amplitude together
with the chosen $K\pi$ S-wave amplitude, reduces Eq.~(\ref{eq48}) to
 \begin{align}
{\mathcal A}_0(k_{\pi},k_{\pi^\prime})=D(k_{\pi},k_{\pi^\prime})+
\tau(M^2_{K\pi})\xi(k_{\pi^\prime})
+\tau(M^2_{K\pi^\prime})\xi(k_{\pi}) \ ,
\label{deq2}
\end{align}
where the interaction between the identical pions is suppressed. The
amplitude given in Eq.~(\ref{deq2}) is a sensible representation of the decay process,
where the $K\pi$ resonant and nonresonant scattering phases are shifted
by the momentum dependent bachelor phase from the three-body rescattering.
The bachelor pion on-mass-shell momentum is given by
\begin{eqnarray}
|{\mathbf
k}_\pi|=\left[\left(\frac{M_D^2+m_\pi^2-M^2_{K\pi^\prime}}{2\,M_D}\right)^2
-m_\pi^2\right]^\frac12 \ ,
\label{kpi}
\end{eqnarray}
with an analogous expression for $|{\mathbf k}_{\pi^\prime}|$. This implies
that each rescattering term in Eq.~(\ref{deq2}) is a function only of $M^2_{K\pi}$ or
$M^2_{K\pi^\prime}$.

\begin{figure}[!htb]
\centering
\includegraphics[scale=0.86]{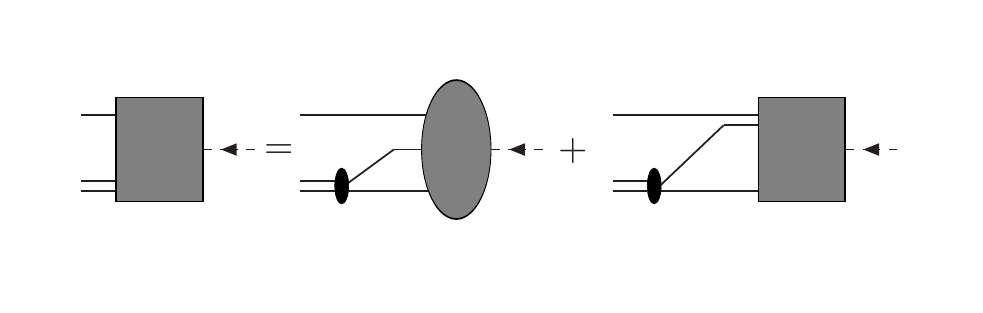}
\vspace{-1cm}
\caption{Diagrammatic representation of the
integral equation for the three-body function
$\tau(M^2_{K\pi^\prime})\xi(k_\pi)$ (gray box). The driving term
contains the partonic amplitude convoluted with the two-body
scattering amplitude (black).}
\label{bsalpeter}
\end{figure}

The resummation of the three-body scattering series results in an inhomogeneous integral
equation for the function $\xi(k)$ of the bachelor momentum,
\begin{align}
\xi(k)=\xi_0(k) +\int\frac{d^4q}{(2\pi)^4}\tau\left((K-q)^2\right)\,S_K(K-k-q)\,S_\pi(q)\,\xi(q),
\label{deq5}
\end{align}
derived from Eq.~(\ref{eq47}) and shown diagrammatically in Fig.~\ref{bsalpeter}. Note
that for convenience, the diagrammatic representation of the integral equation for the
product $\tau\left(M^2_{K\pi^\prime}\right) \xi(k_\pi)$ in presented in the figure.

The driving term
\begin{eqnarray}
\xi_0(k)=\int \frac{d^4q}{(2\pi )^4}S_{\pi}(q)S_K(K-k-q)D(k,q),
\label{deq6}
\end{eqnarray}
carries the partonic decay amplitude to the rescattering process. The second  term in the
rhs of Eq.~(\ref{deq5}) comes from three-body connected diagrams. For example, the lowest
order rescattering term is the connected amplitude given by the third diagram in
Fig.~\ref{ladder}.

Physically, the three-body rescattering in Eq.~(\ref{deq5}) is built by mixing resonances
of the two possible $K\pi$ pairs, and it is a function of the momentum of the bachelor
pion. Therefore, we can say that  the decay amplitude has two contributions: one that is a
smooth function of the momentum of the pions, $D(k_{\pi},k_{\pi^\prime})$, and another
one, $\tau\left(M^2_{K\pi^\prime}\right)\xi(k_\pi)$, that contains the result of the
three-body rescattering, which modulates the $K\pi$ scattering amplitude.

The $K\pi$ S-wave amplitude $\tau$ is an isospin conserving operator acting on the isospin
states $1/2$ and $3/2$. The second and third terms in the rhs of Eq.~(\ref{deq2}) carry
the full effect of the final state interaction through the $K\pi$ scattering amplitude,
considered an operator in isospin space, $\tau$, times a spectator amplitude, $\xi$, that
contains the three-body rescattering contributions. The solution of Eq.~(\ref{deq5}) built
the rescattering series, and the term
$\tau(M^2_{K\pi})\xi(k_{\pi^\prime})+\tau(M^2_{K\pi^\prime})\xi(k_{\pi})$ of the decay
correspond to the sum of the second, third and higher order diagrams depicted in
Fig.~\ref{ladder}. They represent the full hadronic rescattering series of the $K\pi\pi$
system, disregarding three-body irreducible diagrams.

\subsection{Phase and Amplitude Separation}

The S-wave decay amplitude for the $D\to K^-\pi^+\pi^+$ from Eq.~(\ref{deq2}) can be
written as a Bose-symmetrized complex function with respect to the identical pions,
\begin{eqnarray}
{\cal A}_0 = A_0(M^2_{K\pi},M^2_{K\pi^\prime} ) +
A_0(M^2_{K\pi^\prime},M^2_{K\pi } ). \label{boseampl}
\end{eqnarray}
where $A_0$ are complex functions of the two invariant masses squared,  $M^2_{K\pi}= (K-
k_{\pi^\prime})^2$ and $M^2_{K\pi^\prime}= (K- k_{\pi})^2$, which specify the decay
kinematics.

For the $D\to K^-\pi^+\pi^+$ S-wave amplitude in our model, the dependence on the $K\pi$
subsystem mass of $A_0(M^2_{K\pi^\prime},M^2_{K\pi } )$ can be reduced to a complex
function of only one variable $M^2_{K\pi^\prime}$ as
\begin{align}
A_0(M^2_{K\pi^\prime})=a_0(M^2_{K\pi^\prime})e^{i\Phi_0(M^2_{K\pi^\prime})} =\frac{1}{2} \langle K\pi\pi|D\rangle+\langle K\pi\pi|\tau(M^2_{K\pi^\prime})
|\xi(k_{\pi})\rangle,
\label{kpiampl}
\end{align}
where the bachelor pion on-mass-shell momentum is written as a function
$M^2_{K\pi^\prime}$ as given by Eq.~(\ref{kpi}), and $| K\pi\pi\rangle$ represents
the state in isospin space.

\section{FSI Light-Front Dynamics  in Heavy Meson Decay}
\label{sec:FSILF}

The projection onto the light-front (LF) of the four-dimensional field-theoretical  heavy
meson three-particle decay amplitude with FSI, as expressed by Eq.~(\ref{D}), reduces it
to a three-dimensional form. The coupled set of Eqs.~(\ref{eq40}) for the Faddeev
components of the decay amplitude are turned into three-dimensional forms, simplifying
the numerical treatment to solve them.  We follow the LF projection technique of four-dimensional
Bethe-Salpeter like equations as developed in Ref.~\cite{sales00} based on the
quasi-potential approach (QPA). The reduced amplitudes derived using the tools developed
in a series of works \cite{sales00,MarPRD07,MarPoS08,MarFBS08,MarPRD08} and reviewed in
\cite{FredFBS11},  depends only three-dimensional variables, namely, the
kinematical LF momentum  $\underline{k}\equiv(k^+,\vec
k_\perp)$, defined by $k^+=k^0+k^3$ and $\vec k_\perp=\{k_x,k_y\}$. The phase-space integration
is normalized according to $dk^+d^2k_\perp /2(2\pi)^3$.

\subsection{QPA and Decay Amplitude}

The potential in the four-dimensional equation for the three-boson
BSE is given by Eq.~(\ref{presc}) and in terms of the quasi-potential formulation,
the BSE for the transition matrix Eq.~(\ref{eq1}), is substituted  by
\begin{equation}
T=W+W\widetilde{G}_0T\ .  \label{qpa1}
\end{equation}
The quasi-potential $W$ and auxiliary Green's function $(\widetilde{G}_0)$ keep
the dynamical content of the original BSE, when $W$ is the solution of
\begin{equation}
W=V+V\Delta_0W\ , \label{qpa2}
\end{equation}
with $\Delta_0:=G_0-\widetilde{G}_0\label{QP}$. The decay amplitude given by
Eqs.~(\ref{d1}) and (\ref{d2}), can be written in terms of the full three-body T-matrix as
\begin{eqnarray}
{\cal A}=D+T\,G_0\,D \ , \label{qpt0}
\end{eqnarray}
and inserting the QP equation (\ref{qpa1}) in Eq.~(\ref{qpt0}), one has that
\begin{eqnarray}
T\,G_0\,D= W\,G_0\,D+W\,\widetilde G_0\,T\,G_0\,D \ . \label{qpt1}
\end{eqnarray}

The QPA allows to perform a three-dimensional reduction of the four-dimensional
equation (\ref{qpt0}). In particular, the auxiliary Green's function
$\widetilde{G}_0$ can be conveniently chosen to project the four-dimensional
three-body equation (\ref{qpt1}) onto the light-front hypersurface (see \cite{sales00}), and
formally it reads
\begin{equation}
 \widetilde{G}_0:=G_0|~{g_0^{-1}}~ |G_0 \ , \label{agreen}
\end{equation}
where $g_0=|G_0|$ is the free light-front resolvent, including phase-space
factors. The ``bar'' operation on the right or on the left of a four-dimensional
matrix element corresponds to the integration over $k^-=k^0+k^3$, which
eliminates the relative light-front time between
the particles. In our three-particle case, the elimination of the relative
LF time requires an integration  over two independent momenta $k^-$, due to four-momentum conservation,
and we introduce the following operation
\begin{eqnarray}
|A&:=&\int dk_1^-dk_2^-\langle k_1^-k_2^-|A,\nonumber \\
A|&:=&\int dk_1^-dk_2^-A|k_1^-k_2^-\rangle,
\label{bardef3}
\end{eqnarray}
with $A$ being a matrix element of an operator that has matrix elements
function of two independent momenta after the center of mass motion
is factorized.

Explicitly the free three-particle Green's function is given by
\begin{multline}
\langle k_1^-,k_2^- |G_0|k_1^{\prime -},k_2^{\prime -}
\rangle=\frac{-i}{(2\pi)^2}\frac{\delta(k_1^--k_1^{\prime -})}
{\hat{k}_1^+\hat{k}_2^+(K^+-{\hat{k}_1}^+-\hat{k}_2^+)
(k_1^--\hat{k}_{1 on}^-)}\\ \times
\frac{\delta(k_2^--k_2^{\prime -}) }{(k_2^--\hat{k}_{2
on}^-)(K^--k_1^--k_2^--(K-\hat{k}_1-\hat{k}_2)_{on}^-)}~,
\end{multline}
where the hat means operator character and the on-minus-shell momentum $k^-_{ion}=(\vec k^2_\perp+m^2_i)/k^+$.
The on-minus-shell momentum $(K-\hat{k}_1-\hat{k}_2)_{on}^-$ carries the mass of the third
particle $m_3$. By performing the LF projection using Eq.~(\ref{bardef3}), the free LF
Green's function comes as
\begin{eqnarray}
g_0(\underline{k}_1,\underline{k}_2)
=\frac{i\theta(K^+-k_1^+-k_2^+)\theta(k_1^+)\theta(k_2^+)}{k_1^+k_2^+(K^+-{k_1}^+-k_2^+)(K^--k_{1 on}^--k_{2 on}^--(K-k_1-k_2)_{on}^-)}~,\label{propfl3}
\end{eqnarray}

In Refs.~\cite{sales00,FredFBS11} the reader can follow the details of the formal
manipulations within QPA used to project onto the light-front the BSE. Two convenient
operators were introduced in Ref.~\cite{MarPRD08}, which helps to make the notation more
transparent, namely the so-called free light-front reversed operators
\begin{eqnarray}
 \overline{\Pi}_0=G_0|~g_0^{-1}\, ,\,\,  \Pi_0=g_0^{-1}~|G_0 \ , \label{pi0}
\end{eqnarray}
which can only be applied to the right and to left of a three-body four-dimensional
quantity, respectively. These operators also transforms a tridimensional quantity to four dimensional
ones, when acting on the left and on the right of an amplitude dependent on the  kinematical
light-front momenta, respectively. For example, with these operators, we have that the auxiliary Green's
function (\ref{agreen}) is simply written as
\begin{equation}
 \widetilde{G}_0= \overline\Pi_0\, g_0\, \Pi_0.
\end{equation}

Our aim is to obtain the decay amplitude of the heavy meson in three mesons in the final
state, using the three-dimensional projection onto the LF of Eq.~(\ref{qpt1}). By applying
the projection operator $\Pi_0$ in Eq.~(\ref{qpt1}), we get that
\begin{eqnarray}
\Pi_0T\,G_0\,D= \Pi_0W\,G_0\,D+\Pi_0W\,\widetilde G_0\,T\,G_0\,D \ , \label{qpt4}
\end{eqnarray}
which translates to
\begin{eqnarray}
D_{LF}\equiv|G_0TG_0D=|G_0WG_0D+w\,g_0D_{LF}, \label{qpt5}
\end{eqnarray}
after the explicit form given in Eq.~(\ref{pi0}) is used. The function $D_{LF}$ depends
only on the independent kinematical LF momenta of the particles, and the key dynamical
ingredient is the effective LF potential $w=|G_0\,W\,G_0|$ containing the interaction
among the three particles.

\subsection{Effective LF interaction for three-particles}

In order to calculate $w$, we decompose the QP Eq.~(\ref{qpa2}) in three terms, each
given by
\begin{eqnarray}
W_i=V_i+V_i\,\Delta_0\,W~ \label{fdw1}
\end{eqnarray}
with $W$ being the sum over the Faddeev components, i.e., $W=\sum_iW_i$, and
$w=\sum_iw_i= \sum_i|G_0\,W_i\,G_0|$.

The integral equation for the Faddeev component of the quasi-potential is obtained from
the classical form by reintroducing $W$ as a sum of three terms in Eq.~(\ref{fdw1}),
giving
\begin{eqnarray}
 W_i=V_i+V_i\Delta_0(W_i+W_j+W_k) \,
 \end{eqnarray}
which can be rewritten as $(1-V_i\Delta_0)W_i=V_i+V_i\Delta_0(W_j+W_k)$, and
multiplying to the right by  $(1-V_i\Delta_0)^{-1}$, one has that
\begin{eqnarray}
W_i=W_{(2)i}+W_{(2)i}\Delta_0(W_j+W_k), \label{w3}
\end{eqnarray}
where the two-body quasi-potential within the three-body system is
\begin{eqnarray}
 W_{(2)i}=V_{i}+V_{i}\Delta_0W_{(2)i}
 , \label{w2}
\end{eqnarray}
for particle $i$ acting as a spectator.

The solution of Eq.~(\ref{w3}) is obtained in a form of an expansion in powers of $V_i$
where the series for the two-body quasi-potential,
$W_{(2)i}= V_{i}+V_{i}\Delta_0V_i+V_{i}\Delta_0V_{i}\Delta_0V_i+\cdots$,
is used, and terms in $V_i$ collected. The result is
\begin{align}
W_i=V_i+V_i\Delta_0(V_i+V_j+V_k)
+ V_i\Delta_0(V_i+V_j+V_k)\Delta_0(V_i+V_j+V_k)+\dots \, .
\label{Wi}
\end{align}
The leading order (LO) and next-to-leading-order (NLO) terms, the first and second power
in the interaction $V_i$, are given by $W^{LO}_i=V_i$ and by
$W^{NLO}_i=V_i+V_i\Delta_0(V_i+V_j+V_k)$, respectively. Therefore, the Faddeev components
of LF effective potential in LO and NLO are written in terms of the above expansion as
\begin{align}
w^{LO}_i&={g_0^{-1}}~ |G_0V_iG_0|~g_0^{-1}\ , \\
w^{NLO}_i&=w^{LO}_i
+{g_0^{-1}}|G_0V_i\Delta_0(V_i+V_j+V_k)G_0|~g_0^{-1} ~.
\end{align}
The effective interactions $w_i$ builds the dynamical equation
for the decay amplitude $D_{LF}$, Eq.~(\ref{qpt5}), and the leading order
calculation corresponds to a truncation at the valence states, which will be
used in the next to built model for the heavy meson decay. We should note that the NLO interaction
includes induced light-front three-body forces, namely terms like $g_0^{-1}|G_0\,V_i\Delta_0\,V_j\,G_0|~g_0^{-1}$,
and already pointed out in \cite{KarFBS09}.

\subsection{LF Faddeev equations for $D_{LF}$}

The LF projected decay amplitude solution of Eq.~(\ref{qpt5}) is decomposed
in a sum $D_{LF}=\sum_i D^i_{LF} $, where the Faddeev components are
\begin{eqnarray}
D^i_{LF}=|G_0W_iG_0D+w_i\,g_0D_{LF}.
\label{qpt6}
\end{eqnarray}
The standard manipulation leads to
\begin{eqnarray}
D^i_{LF}=d^i_{LF,0}+t_i\,g_0\left(D^j_{LF}+ D^k_{LF}\right),
\end{eqnarray}
where $d^i_{LF,0}=(1-w_ig_0)^{-1}|G_0W_iG_0D$ and the reduced LF transition matrix $t_i$ is the solution of
$t_i=w_i-w_ig_0t_i$. Assuming, that the partonic amplitude $D$ is weakly dependent
in $k^-$ and  the main dependence on $k^-$ in the integrand comes from the free propagator and $W_i$,
we can write that
\begin{align}
d^i_{LF,0}=(1-w_ig_0)^{-1}|G_0W_iG_0|D
=(1-w_ig_0)^{-1}g_0w_ig_0D
=t_ig_0D,
\label{qpt7}
\end{align}
which will be exactly valid if $D$ is constant, as in our numerical application. The LF
Faddev equation for the component of the vertex simply becomes
\begin{eqnarray}
D^i_{LF}=t_ig_0D+t_i\,g_0\left(D^j_{LF}+ D^k_{LF}\right), \label{qpt8}
\end{eqnarray}
and in next the model for $t_i$ is considered.

The LF front model for the two-body scattering amplitude comes from Eq.~(\ref{eq42}),
using the relation $t_i=\overline\Pi_0T_i\Pi_0$:
\begin{multline}
\langle
\underline{k}'_j,\underline{k}'_k|g_0t_ig_0|\underline{k}_j,\underline{k}_k\rangle =
\langle \underline{k}'_j,\underline{k}'_k|\, |G_0T_{i}G_0|\,|\underline{k}_j,\underline{k}_k\rangle=
\\ =(2\pi)^4\int dk_j^{\prime-}dk_k^{\prime-}
\int dk_j^-dk_k^-
\int dq_j^-dq_k^-
\int dq_j^{\prime-}dq_k^{\prime-}
 \,S^{-1}_i(q_i)\,\delta(q^\prime_i-q_i)\langle k_j^{\prime-}k_k^{\prime-}|G_0| q_j^{\prime-}q_k^{\prime-}\rangle\, \\ \times
 \tau_i(s_{i})\,
\langle q_j^{-}q_k^- |\ G_0| k_j^{-}k_k^{-} \rangle
 \\
=2\delta(\underline K-\underline k^\prime_j-\underline k^\prime_k-\underline k_i)
\int dk_j^{\prime-} dk_j^-dk_i^-i^6 \,S_j(k^\prime_j)S_k(K-k^\prime_j-k_i)\\ \times  \tau_i((K-k_i)^2)
S_i(k_i)S_j(k_j)S_k(K-k_j-k_i).\label{eq42t}
\end{multline}

Performing the Cauchy integration in each variable $k_j^{\prime-}$, $k_j^-$ and $k_i^-$, and
given that $\tau_i((K-k_i)^2)$ is analytical in the lower-half of the
$k^-_i$ complex-plane, the result is
\begin{eqnarray}
\langle \underline{k}'_j,\underline{k}'_k|t_i|\underline{k}_j,\underline{k}_k\rangle =
2(2\pi)^3k^+_i\delta(\underline k^\prime_i-\underline k_i)
\tau_i\left(M^2_{jk}\right)\ , \label{eq43t}
\end{eqnarray}
where $M^2_{jk}=(K-k_{ion})^2$. Owing to the separable form of the two-body amplitude, the
Faddeev component of the decay amplitude separates as
\begin{eqnarray}
D^i_{LF}(\underline k_j,\underline k_k)=\tau_i\left(M_{jk}^2\right)\xi^i(\underline k_i), \label{qpt9}
\end{eqnarray}
as also happens for the four-dimensional case shown in Eq.~(\ref{eq43}).

The integral equation for the reduced decay amplitude, $\xi^i(\underline k_i)$ becomes
\begin{small}
\begin{multline}
\xi^i(\underline k_i)=\xi^i_0(\underline k_i)+ \frac{i}{2(2\pi)^3}\int_0^{K^+-k^+_i}
\frac{dq^+_j }{q^+_j(K^+-k^+_i-q^+_j)}\int d^2q_{j\perp}\frac{\tau_j\left((K-q_{jon}
)^2\right)\xi^j(\underline q_j)}{K^--k^-_{ion}-q^-_{jon}-(K-k_i-q_j)^-_{on}+i\varepsilon}
 \\
+\frac{i}{2(2\pi)^3}\int_0^{K^+-k^+_i} \frac{dq^+_k }{q^+_k(K^+-k^+_i-q^+_k)}\int
d^2q_{k\perp}\frac{\tau_k\left((K-q_{kon})^2\right)\xi^k(\underline q_k)}{K^--k^-_{ion}
-(K-k_i-q_k)^-_{on}-q^-_{kon}+i\varepsilon},
\label{qpt10}
\end{multline}
\end{small}
where
\begin{small}
\begin{eqnarray}
\xi^i_0(\underline k_i)=\frac{i}{2(2\pi)^3}\int_0^{K^+-k^+_i} \frac{dq^+_j }{q^+_j(K^+-k^+_i-q^+_j)}\int d^2q_{j\perp}
\frac{D(\underline q_j;\underline k_i)}{K^--k^-_{ion}-q^-_{jon}-(K-k_i-q_j)^-_{on}+i\varepsilon}
 . \label{qpt11}
  \end{eqnarray}
 \end{small}

Rewriting Eqs.~(\ref{qpt10}) and (\ref{qpt11}) in terms of momentum fractions, one gets
\begin{eqnarray}
\xi^i(y,\vec k_\perp) = \xi^i_0(y,\vec k_\perp)+ \frac{i}{2(2\pi)^3}\int_0^{1-y} \frac{dx }{x(1-x-y)}\int d^2q_{\perp}
\left[\frac{\tau_j\left(M^2_{ik}(x,q_\perp)\right)\xi^j(x,\vec q_\perp)}{M^2-M_{0}^2(x,\vec q_\perp;y,\vec k_\perp)+i\varepsilon}
+ (j\leftrightarrow k)
\right]
, \label{qpt12}
  \end{eqnarray}
where $M^2=K^\mu K_\mu$,  $y=k^+_i/K^+$, $x=q^+_j/K^+$ or $x=q^+_k/K^+$  in the first or second integral in the right-hand side of the equation.
The free three-body squared  mass is
\begin{eqnarray}
 M_0^2(x,\vec q_\perp;y,\vec k_\perp)=\frac{k_\perp^2+m^2_i}{y}
 +\frac{q_\perp^2+m^2_j}{x} + \frac{(\vec k_\perp+\vec q_\perp)^2+m^2_k}{1-x-y}.
\label{qpt14}
\end{eqnarray}
The argument of the two-body amplitude $\tau_j\left(M^2_{ik}(x,q_\perp)\right)$ should be
understood as
\begin{eqnarray}
 M_{ik}^2(x,q_\bot) = (1-x)\left(M^2 - \frac{q_\bot^2 +
m_j^2}{x}\right) - q_\bot^2 \ .
\end{eqnarray}
The driven term in Eq.~(\ref{qpt12}) is rewritten as
\begin{eqnarray}
\xi^i_0(y,\vec k_\perp)= \frac{i}{2(2\pi)^3} \int_0^{1-y} \frac{dx}{x(1-y-x)}\int d^2q_{\perp}
\frac{D(x,\vec q_\perp;y,\vec k_\perp)}{M^2-M_0^2(x,\vec q_\perp;y,\vec k_\perp)+i\varepsilon}
 . \label{qpt13}
\end{eqnarray}
The LF model for the three-body heavy meson decay modeled by Eqs.~(\ref{qpt9})
and (\ref{qpt12}) assumes the dominance of the valence state in the intermediate state
propagations and the $s$-channel description of the two-meson amplitude.
To be complete, the LF counterpart of the decay amplitude in Eq.~(\ref{eq48}) is
\begin{eqnarray}
{\mathcal A}_0=D+\sum_\alpha\tau(s_\alpha)\xi^\alpha(y,\vec k_\perp) \ ,
\label{eq48lf}
\end{eqnarray}
where $s_\alpha=(K-k_{\alpha on})^2$  and the partonic function $D$ is a function on the
momentum of the on-mass-shell particles in the decay channel. We concluded the general
formalism for the calculation of the heavy meson decay amplitude in three spinless mesons.
For the $D^+\to K^-\pi^+\pi^+$ process Eq.~(\ref{eq48lf}) reduces to
\begin{align}
{\mathcal A}_0(k_{\pi},k_{\pi^\prime})=D(k_{\pi},k_{\pi^\prime}) +\tau(M^2_{K\pi})\xi(\underline k_{\pi^\prime})+\tau(M^2_{K\pi^\prime})\xi(\underline
k_{\pi})\, ,
\label{deq2lf}
\end{align}
where as we have assumed, also in the four-dimensional case, see Eq.~(\ref{deq2}), the
interaction between the identical pions is suppressed. In order to keep the rotation
invariance of the calculation, the $z-$direction is chosen transverse to the decay plane
in the rest frame of the $D^\pm$ meson. This choice makes optimal use of the kinematical
nature of the rotation in the transverse plane, adopted as the plane where the momentum
of each meson in the final state are.

\section{LF model for $D^+\to K^-\pi^+\pi^+$ decay}
\label{LFMDDecay}

The light-front model for the $D^+\to K^-\pi^+\pi^+$ decay with FSI is given by
the inhomogeneous integral equation for the bachelor meson amplitude (\ref{qpt12}), with
the driven term (\ref{qpt13}), and full decay amplitude written in Eq.~(\ref{deq2lf}).
Besides the partonic amplitude, which defines the driven term for the bachelor amplitude,
the two-meson scattering amplitude is the input for the calculations. We disregard the
$\pi\pi$ interaction in isospin $1$ charged states, and consider only the neutral channels
$K\pi$ states. The isospin states for $K^\mp\pi^\pm$ are $I_{K\pi}=1/2$ and
$I_{K\pi}=3/2$, the parametrization of the S-waves amplitudes given in
Sec.~\ref{sec:kpiampl}. The dominant $K\pi$ amplitude is the resonant $I_{K\pi}=1/2$ one
below $K^*_0(1430)$, but above it the $I_{K\pi}=3/2$ amplitude has a comparable
contribution for the scattering \cite{LASS}. Therefore, to explore the available
phase-space for the $D$ decay above  $K^*_0(1430)$, one has to consider not only
$I_{K\pi}=1/2$ but also $I_{K\pi}=3/2$. Indeed, below $K^*_0(1430)$ the calculations
were previously performed in Ref.~\cite{MagPRD11}. It was also included the $I_{K\pi}=3/2$
interaction in the $D$ decay amplitude up to two-loops in Ref.~\cite{MagPoS12}.

In this section, we present a isospin conserving light-front model, including interaction
in both $K\pi$ isospin states, and perform calculations up to three-loops, in order to
check the convergence of the results. The possible total isospin states $(I_T)$
are $5/2$ and $3/2$ with $I_T^z=\pm 3/2$. The bachelor amplitude, solution
of Eq.~(\ref{qpt12}), carries  the total isospin index, and the interacting pair isospin,
namely $\xi^{I_T^z}_{I_T,I_{K\pi}}(y,k_\bot)$, we keep for convenience, the isospin
projection. We restrict our calculations only to s-wave states and the bachelor amplitude
depends only on $|\vec k_\perp|\equiv k_\perp$. The partonic decay amplitude has now to be
projected on two $K\pi$ isospin states, and total isospin, i.e.,
\begin{align}
\left|D\right> = \sum_{I_T,I_{K\pi}} \alpha^{I_T^z}_{I_T,I_{K\pi}}
\left|I_T,I_{K\pi},I_T^z\right> + \sum_{I_T,I_{K\pi^\prime}} \alpha^{I_T^z}_{I_T,I_{K\pi^\prime}}
\left|I_T,I_{K\pi^\prime},I_T^z\right>.
\label{initial}
\end{align}

The projected LF inhomogeneous integral equations  for the bachelor amplitudes $\xi^{I_T^z}_{I_T,I_{K\pi}}$
built from Eq.~(\ref{qpt12}), are given by a set of isospin coupled systems, with the
driven term weighted by the partonic amplitude (\ref{initial}), and written as
\begin{multline}
\xi^{I_T^z}_{I_T,I_{K\pi}}(y,k_\bot)=\left<I_T,I_{K\pi},I_T^z|D\right>
\xi_0(y,k_\bot)
+\frac{i}{2 }\sum_{I_{K\pi^\prime}}R^{I_T^z}_{I_T,I_{K\pi},I_{K\pi^\prime}}
\int_0^{1-y}\frac{dx}{x(1-y-x)}\int_0^\infty \frac{dq_\bot  }{(2\pi)^3} \\  \times
 K_{I_{K\pi^\prime}}(y,k_\bot;x,q_\bot)\,\xi^{I_T^z}_{I_T,I_{K\pi^\prime}}(x,q_\bot), \label{quilfisos}
\end{multline}
where the kernel carrying the isospin of the interacting pair is
\begin{eqnarray}
K_{I_{K\pi^\prime}}(y,k_\bot;x,q_\bot)=\int_0^{2\pi} d\theta \,\,
{q_\bot\,\tau_{I_{K\pi^\prime}}\left(M_{K\pi^\prime}^2(x,q_\bot)\right)\over M_D^2-M_{0,K\pi\pi}^2(x,q_\bot,y,k_\bot)+i\varepsilon}.
\label{kisos}
\end{eqnarray}
The isospin recoupling coefficients in Eq.~(\ref{quilfisos}) are
$R^{I_T^z}_{I_T,I_{K\pi},I_{K\pi^\prime}}=\left<I_T,I_{K\pi},I_T^z|I_T,I_{
K\pi^\prime},I_T^z\right>$. The free squared mass of the $K\pi\pi$  system is
\begin{eqnarray}
M_{0,K\pi\pi}^2(x,q_\bot,y,k_\bot) = \frac{k_\bot^2 + m_\pi^2}{y}
+ \frac{q_\bot^2 + m_\pi^2}{x}
+ \frac{q_\bot^2 + k_\bot^2 + 2q_\bot k_\bot \cos\theta+ m_K^2}{1-x-y},
\end{eqnarray}
and the squared-mass of the virtual $K\pi$ system is
\begin{eqnarray}
M_{K\pi}^2(z,p_\bot) =(1-z)\left(M_D^2 - \frac{p_\bot^2 +
m_\pi^2}{z}\right) - p_\bot^2. \,\,\,
\label{mkpi2}
\end{eqnarray}

The driving term is regularized by one subtraction, at the scale $\mu$, and one finite
subtraction constant $\lambda(\mu^2)$, and is written as
\begin{small}
\begin{eqnarray}
\xi_0(y,k_\bot)=\lambda(\mu^2)+\frac{i}{2 }\int_0^1\frac{dx}{x(1-x)}\int_0^{2\pi}
d\theta\int_0^\infty \frac{dq_\bot q_\bot}{(2\pi)^3} \left[\frac{1}{M_{K\pi}^2(y,k_\bot)-M_{0,K\pi}^2(x,q_\bot)
+ i\varepsilon} -\frac{1}{\mu^2-M_{ 0,K\pi}^2(x,q_\bot)}\right],
\label{xi0}
\end{eqnarray}
\end{small}
where the free squared-mass of the virtual $K\pi$ system in the driven term is
\begin{eqnarray}
M_{ 0,K\pi}^2(x,q_\bot) = \frac{q_\bot^2 + m_\pi^2}{x} +
\frac{q_\bot^2 + m_K^2}{1-x}.
\end{eqnarray}
Performing the angular and radial integrations one gets that
\begin{eqnarray}
\xi_0(y,k_\bot)= \lambda(\mu^2)+ \frac{i}{4}\int_0^1 \frac{dx}{(2\pi)^2} \ln \frac{\Lambda_1}{\Lambda_2},
\label{xi01}
\end{eqnarray}
where
\begin{align}
\Lambda_1&=(1-x)(xM_{K\pi}^2(y,k_\bot)-m_\pi^2+ix\varepsilon)-xm_K^2 \ ,\\
\Lambda_2&=(1-x)(x\mu^2-m_\pi^2)-xm_K^2 \ .
 \end{align}
The light-front $D$ decay model with isospin dependence on the $K\pi$ pair will be explored further
in two situations: $i)$ only $K\pi$ s-wave interaction in the resonant $I=1/2$ state
(single-channel model); and $ii)$ $K\pi$
s-wave interaction in $I=1/2$ and 3/2 (coupled-channel model). In  both cases we disregard the
pion-pion interaction in $I=2$ states.

\subsection{Phase and Amplitude Separation}

The full $D\to K^-\pi^+\pi^+$ S-wave decay amplitude  from the solution of
Eq.~(\ref{deq2}) is symmetrized in respect to the identical pions as given in
Eq.~(\ref{boseampl}), and written as
\begin{eqnarray}
{\cal A}_0 = A_0(M^2_{K\pi^\prime} ) +
A_0(M^2_{K\pi } ). \label{boseamplf}
\end{eqnarray}
as each amplitude depends only on the Mandelstam \mbox{s-variable} of each $K\pi$
subsystem. In detail, each amplitude in Eq.~(\ref{boseamplf}) has the bachelor amplitude
and $K\pi$ scattering amplitude, i.e.,
\begin{eqnarray}
A_0(M^2_{K\pi^\prime})
&=&\sum_{I_T,I_{K\pi^\prime},I_T^z}\left<K^-\pi^+\pi^+\right|
\left.I_T,I_{K\pi^\prime},I_T^z\right> \left[\frac{1}{2}\left<I_T,I_{K\pi^\prime},I_T^z\right|
\left.D\right>+\tau_{I_{K\pi }}(M^2_{K\pi^\prime})\xi^{I_T^z}_{I_T,I_{K\pi^\prime}}
(k_{\pi})\right] \nonumber \\ &=&a_0(M^2_{K\pi^\prime})e^{i
\Phi_0(M^2_{K\pi^\prime})}
\ , \label{kpiamplf}
\end{eqnarray}
where the projection onto the final $K\pi\pi$ isospin state is performed. In the bachelor
amplitude the pion momentum is on-mass-shell and, due to the total momentum conservation,
its modulus is defined by Eq.~(\ref{kpi}) as a function of $M^2_{K\pi^\prime}$.

\section{Perturbative solutions}
\label{pertsol}

The perturbative solution of the integral equations for the bachelor amplitudes
(\ref{quilfisos}) in the different isospin channels, is found by iteration starting from
the driving term. The terms in the perturbative series are obtained by loop integrations.
The bachelor amplitude found from the driving term corresponds to a one-loop calculation.
In Ref.~\cite{MagPRD11}, a calculation up to two-loops were performed. For the
single-channel model, we calculate up to three-loops to check the convergence of the
perturbative series. For the coupled-channel, where the total isospin states $I=3/2$ can
be formed either by coupling $I_{K\pi}=1/2$ or $I_{K\pi}=3/2$, we also perform
calculations up to three-loops. Also the $I_T=5/2$ is considered, where the only
contribution from the $K\pi$ interaction happens in the isospin 3/2 states. Indeed such
contributions to the $K\pi\pi$ phase are marginal.

\subsection{Interaction in $I_{K\pi}=1/2$ state}

Our aim in the single channel example is to solve numerically the light-front
Eq.~(\ref{quilfisos}) for the bachelor amplitude when we consider only $K\pi$ interaction
in the resonant isospin 1/2 states. In this case, Eq.~(\ref{quilfisos}) reduces to
\begin{eqnarray}
\xi^{\mbox{\tiny{3/2}}}_{\mbox{\tiny{3/2,1/2}}}(y,k_\bot)= \frac{1}{6}\sqrt{\frac
{2}{3}}\xi_0(y,k_\bot) -
\frac{i}{3 }\int_0^{1-y}\frac{dx}{x(1-y-x)} \int_0^\infty \frac{dq_\bot}{(2\pi)^3}
K_{\mbox{ \tiny{1/2}}}(y,k_\bot;x,q_\bot)\,\xi^{\mbox{
\tiny{3/2}}}_{\mbox{\tiny{3/2,1/2}}}(x,q_\bot),
\label{eqint}
\end{eqnarray}
where the driving term is computed by considering only $\alpha^{3/2}_{3/2,1/2}$ in
Eq.~(\ref{initial}) nonvanishing and equal to unity, and the partonic decay
amplitude is assumed momentum independent.

The perturbative solution of the integral equation (\ref{eqint}) up to three-loops is
given by
\begin{multline}
\xi^{\mbox{\tiny{3/2}}}_{\mbox{\tiny{3/2,1/2}}}(y,k_\bot) =
\frac{1}{6}\sqrt{\frac{2}{3}}\xi_0(y,k_\bot) -
\frac{i}{3 }\left(\frac{1}{6}\sqrt{\frac{2}{3}}\right)
\int_0^\infty \frac{dq_\bot}{(2\pi)^3}\int_0^{1-y}dx\,K_{\mbox{\tiny 1/2}}(y,k_\bot;x,q_\bot)
\,\xi_0(x,q_\bot)+\\ -
\frac{1}{9}\left(\frac{1}{6}\sqrt{\frac{2}{3}}
\right)\int_0^\infty \frac{dq_\bot}{(2\pi)^3}\int_0^{1-y}dx\,
K_{\mbox{\tiny 1/2}}(y,k_\bot;x,q_\bot)
\int_0^\infty \frac{dq_\bot'}{(2\pi)^3}\int_0^{1-x}dx'\,K_{\mbox{\tiny
1/2}}(x,q_\bot;x',q_\bot')\,\xi_0(x',
q_\bot')+\cdots
\label{itersol-single}
\end{multline}
where $K_{\mbox{\tiny 1/2}}(x,q_\bot,y,k_\bot)$ is defined by Eq.~(\ref{kisos}).

\begin{figure}[!thb]
\centering
\includegraphics[scale=0.5]{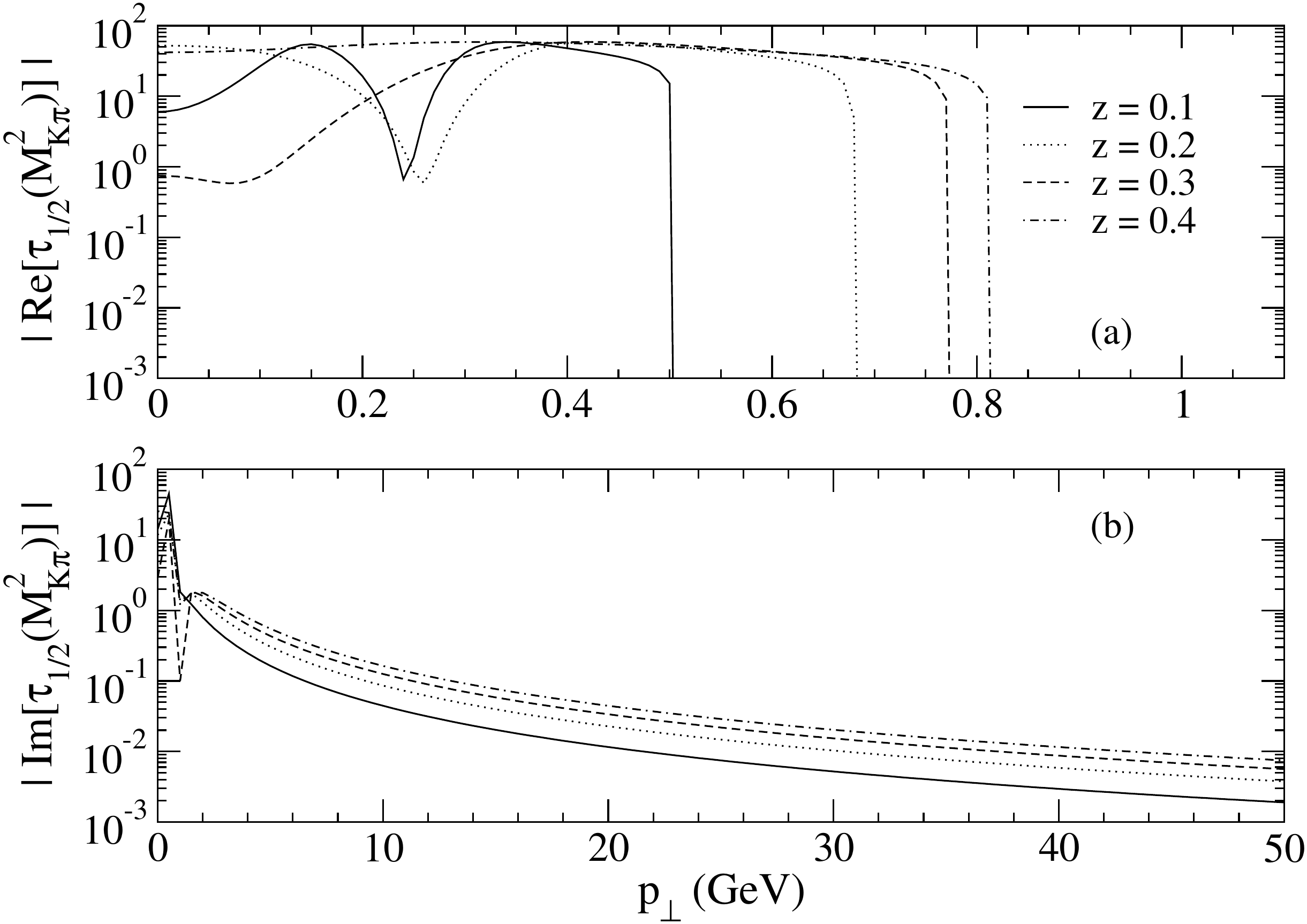}
\caption{Real and imaginary parts of the $I=1/2$ S-wave $K\pi$ amplitude as a function of
$p_\bot$ for some $z$ values in the unphysical mass region.  The $M^2_{K\pi}$ value is
related to $p_\bot$ and $z$ through Eq.~(\ref{mkpi2}).}
\label{tau}
\end{figure}

For the numerical calculation  of the bachelor amplitude up to three-loops in
Eq.~(\ref{itersol-single}), we introduce a momentum cut-off, $\Lambda=2$~GeV for numerical
convenience. Note that the imaginary part of the  s-wave $K\pi$ amplitude
from Eqs.~(\ref{tau12}) and (\ref{skpi12}) in the unphysical region, as plotted in
Fig.~\ref{tau}, goes fast enough to zero for large momentum and shows that the momentum
loop integrals in the perturbative calculation in Eq.~(\ref{itersol-single}) are finite. In
the figure, we plot the real and imaginary parts of the  $K\pi$ amplitude as a
function of $z$ and $p_\bot$, which are the arguments of the squared mass of the
interacting virtual $K\pi$ system given in Eq.~(\ref{mkpi2}), and corresponds to $x$ and
$q_\bot$ in the kernel of Eq.~(\ref{itersol-single}), respectively.

The analytic continuation of the s-wave isospin $1/2$ $K\pi$ scattering amplitude to the
unphysical region  of the $K\pi$ amplitude, i.e., for $M^2_{K\pi}< (m_K+m_\pi)^2$, is
chosen as the imaginary part of $\tau_{1/2}$, with the effective range in
Eq.~(\ref{effexp}) turned off. For the isospin $3/2$ case, also the effective range in
Eq.~(\ref{tau32}) is disregarded in the unphysical region in order to avoid bound states
poles in the S-matrix.  Note that in the kernel of the integral equations only the
imaginary part of the $K\pi$ amplitude is used in the unphysical region, which corresponds
to a real scattering amplitude, as it should be.

In our calculations of Eq.~(\ref{itersol-single}), we have considered finite values of for
$\varepsilon$ in the meson propagators, we use 0.2 and 0.3 GeV, which induces absorption
and mimics coupling to other decay channels. The subtraction constant in the driving term
is chosen for $\mu^2=0$ to be $\lambda (0)=0.12+i 0.06$, which matches the driving term
computed in Ref.~\cite{MagPRD11}. We test the change in the subtraction parameter, by
keeping $\lambda(\mu^2)$ fixed to $\lambda(0)$, while moving~$\mu^2$.

\begin{figure}[!htb]
\centering
\includegraphics[scale=0.5]{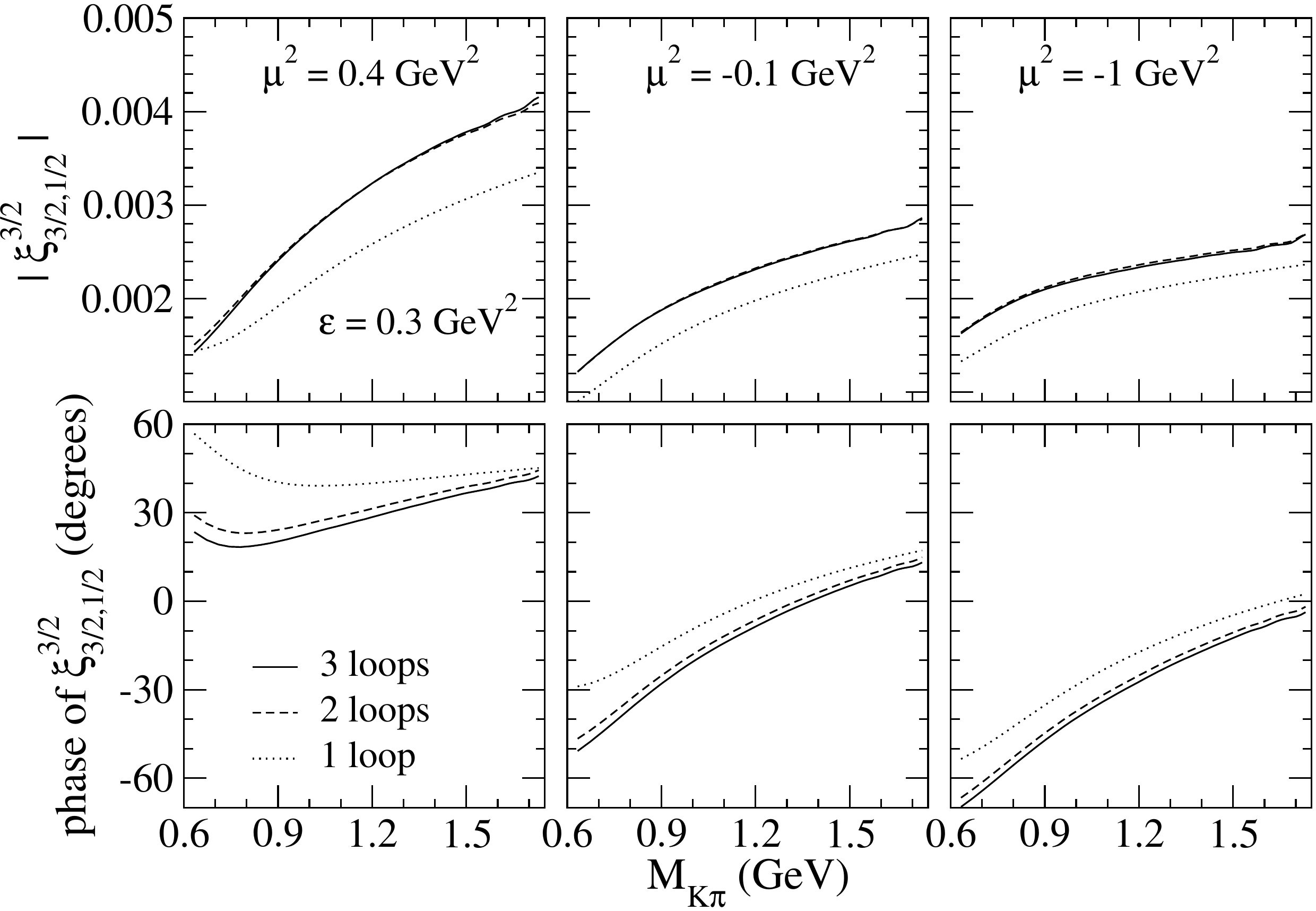}
\caption{Modulus and phase of $\xi^{\mbox{\tiny{3/2}}}_{\mbox{\tiny{3/2,1/2}}}$
for $\varepsilon=0.3$ GeV$^2$ and different $\mu^2$ values.}
\label{qsisingle}
\end{figure}

In Fig.~\ref{qsisingle}, we study the convergence of the loop  expansion for the phase
and amplitude of the bachelor function up to three-loops. We choose
$\mu^2=(0.4,-0.1,-1)$~GeV$^2$ with $\varepsilon=0.3$ GeV$^2$. The values of $|\mu|$ are
chosen within the hadronic scale between $\sim$  0.3 to 1 GeV, spanning values of $\mu^2$
above and below zero in order to verify the sensitivity of the bachelor function.
Irrespectively to the value of $\mu^2$, the 2-loops solution is good enough and can be
used to compute the bachelor amplitude. However, the phase can be either positive or
negative, but it increases with $M^2_{K\pi}$. For $\mu^2=-0.4$~GeV$^2$ and $\mu^2=-1$~
GeV$^2$, the phase difference between the threshold and the maximum for the mass of the
$K\pi$ system, the phase shows a quite large variation of about $60^o$. The modulus
increases with $M^2_{K\pi}$ for all $\mu^2$ values.

\subsection{Interaction in $I_{K\pi}=1/2$ and 3/2 states}

The inclusion of the two possible isospin channels for the $K\pi$ interacting system,
namely, $1/2$ and $3/2$, results in a coupled set of inhomogeneous integral equations from
Eq.~(\ref{quilfisos}) for $I_T=3/2$,
which reads
\begin{multline}
\xi^{\mbox{\tiny{3/2}}}_{\mbox{\tiny{3/2,1/2}}}(y,k_\bot)=A_w\,\xi_0(y,k_\bot)
+\frac{iR^{\mbox{\tiny{3/2}}}_{\mbox{\tiny{3/2,1/2,1/2}}}}{2}\int_0^{1-y
}\frac{dx}{x(1-y-x)}\int_0^\infty \frac{dq_\bot}{(2\pi)^3} K_{\mbox{\tiny 1/2}}(y,k_\bot;x,q_\bot)\,\xi^{\mbox{\tiny{3/2}}
}_{\mbox{\tiny{3/2,1/2}}}(x,q_\bot) + \\
+\frac{iR^{\mbox{\tiny{3/2}}}_{\mbox{\tiny{3/2,1/2,3/2}}}}{2(2\pi)^3}
\int_0^{1-y}\frac{dx}{x(1-y-x)}\int_0^\infty \frac{dq_\bot}{(2\pi)^3}\,
K_{\mbox{\tiny 3/2}}(y,k_\bot;x,q_\bot)\,\xi^{\mbox{\tiny{3/2}}
}_{\mbox{\tiny{3/2,3/2}}}(x,q_\bot),
\label{eqint-coupled1}
\end{multline}
\begin{multline}
\xi^{\mbox{\tiny{3/2}}}_{\mbox{\tiny{3/2,3/2}}}(y,k_\bot)=B_w\, \xi_0(y,k_\bot)
+\frac{iR^{\mbox{\tiny{3/2}}}_{\mbox{\tiny{3/2,3/2,1/2}}}}{2}\int_0^{1-y
}\frac{dx}{x(1-y-x)}\int_0^\infty \frac{dq_\bot}{(2\pi)^3}\,
K_{\mbox{\tiny 1/2}}(y,k_\bot;x,q_\bot)\,\xi^{\mbox{\tiny{3/2}}
}_{\mbox{\tiny{3/2,1/2}}}(x,q_\bot)   + \\
+\frac{iR^{\mbox{\tiny{3/2}}}_{\mbox{\tiny{3/2,3/2,3/2}}}}{2}
\int_0^{1-y}\frac{dx}{x(1-y-x)}\int_0^\infty \frac{dq_\bot}{(2\pi)^3}\,K_{\mbox{\tiny 3/2}}(y,k_\bot;x,q_\bot)\,\xi^{\mbox{\tiny{3/2}}
}_{\mbox{\tiny{3/2,3/2}}}(x,q_\bot) .
\label{eqint-coupled2}
\end{multline}
For $I_T=5/2$ Eq.~(\ref{quilfisos}) is single channel and interaction only in
$I_{K\pi}=$3/2 is possible. In this case, the inhomogeneous equation for the bachelor
amplitude is
\begin{multline}
\xi^{\mbox{\tiny{3/2}}}_{\mbox{\tiny{5/2,3/2}}}(y,k_\bot)=C_w\,\xi_0(y,k_\bot)
+\frac{iR^{\mbox{\tiny{3/2}}}_{\mbox{\tiny{5/2,3/2,3/2}}}}{2 }\int_0^{1-y
}\frac{dx}{x(1-y-x)}\int_0^\infty \frac{dq_\bot}{(2\pi)^3}\,K_{\mbox{\tiny 3/2}}(y,k_\bot;x,q_\bot)
\, \xi^{\mbox{\tiny{3/2}}}_{\mbox{\tiny{5/2,3/2}}}(x,q_\bot),
\label{eqint-coupled3}
\end{multline}
where the weights $A_w$, $B_w$ and $C_w$ appearing in the driving terms are computed from the initial
distribution of isospin states from the partonic amplitude (\ref{initial}).

The weights in the driven terms of Eqs.~(\ref{eqint-coupled2}) and (\ref{eqint-coupled3})
are computed from the overlap of isospin state with the initial isospin distribution of
the decay from the partonic amplitude,
\begin{eqnarray}
\label{awbwcw}
&& A_w=\frac{1}{2}\left<I_T=3/2,I_{K\pi}=1/2,I_T^z=3/2\right|\left.D\right>,\qquad \\
&& B_w=\frac{1}{2}\left<I_T=3/2,I_{K\pi}=3/2,I_T^z=3/2\right|\left.D\right>,\qquad \\
&& C_w=\frac{1}{2}\left<I_T=5/2,I_{K\pi}=3/2,I_T^z=3/2\right|\left.D\right>,\qquad
\end{eqnarray}
and, evaluating in details the isospin coefficients, one gets
\begin{eqnarray}
A_w &=&
\alpha^{\mbox{\tiny{3/2}}}_{\mbox{\tiny{3/2,1/2}}}
(1+R^{\mbox{\tiny{3/2}}}_{\mbox{\tiny{3/2,1/2,1/2}}})
+ \alpha^{\mbox{\tiny{3/2}}}_{\mbox{\tiny{3/2,3/2}}}
R^{\mbox{\tiny{3/2}}}_{\mbox{\tiny{3/2,1/2,3/2}}}, \qquad\\
B_w &=& \alpha^{\mbox{\tiny{3/2}}}_{\mbox{\tiny{3/2,3/2}}}
(1+R^{\mbox{\tiny{3/2}}}_{\mbox{\tiny{3/2,3/2,3/2}}})
+ \alpha^{\mbox{\tiny{3/2}}}_{\mbox{\tiny{3/2,1/2}}}
R^{\mbox{\tiny{3/2}}}_{\mbox{\tiny{3/2,3/2,1/2}}}, \qquad\\
C_w &=& \alpha^{\mbox{\tiny{3/2}}}_{\mbox{\tiny{5/2,3/2}}}
(1+R^{\mbox{\tiny{3/2}}}_{\mbox{\tiny{5/2,3/2,3/2}}}).
\end{eqnarray}
The coefficients $\alpha$ appearing above comes from the initial decay amplitude (\ref{initial}),
and now we define them in terms of the parameters $W_i$ ($i=1,2,3$), such that
\begin{eqnarray}
\alpha^{\mbox{\tiny{3/2}}}_{\mbox{\tiny{3/2,1/2}}}&=&\frac{W_1}{2}\,
C^{\mbox{\tiny{1/2  1   3/2}}}_{\mbox{\tiny{1/2  1   3/2}}} \,
C^{\mbox{\tiny{1   1/2  1/2}}}_{\mbox{\tiny{1  -1/2  1/2}}}\, , \\
\alpha^{\mbox{\tiny{3/2}}}_{\mbox{\tiny{3/2,3/2}}}&=&\frac{W_2}{2}\,
C^{\mbox{\tiny{3/2  1   3/2}}}_{\mbox{\tiny{1/2   1   3/2}}}\,
C^{\mbox{\tiny{1   1/2  3/2}}}_{\mbox{\tiny{1   -1/2  1/2}}}\, , \\
\alpha^{\mbox{\tiny{3/2}}}_{\mbox{\tiny{5/2,3/2}}}&=&\frac{W_3}{2}\,
C^{\mbox{\tiny{3/2  1  5/2}}}_{\mbox{\tiny{1/2 1  3/2}}}\,
C^{\mbox{\tiny{1  1/2  3/2}}}_{\mbox{\tiny{1 -1/2  1/2}}}\, ,
\label{alpha1}
\end{eqnarray}
which in the particular case of $\left|D\right>=\left|K^-\pi^+\pi^+\right>$ one has that
$W_1=W_2=W_3=1$.

To be complete, the respective Clebsch-Gordan and recoupling
coefficients necessary for all computations are
$\,C^{\mbox{\tiny{1/2  1   3/2}}}_{\mbox{\tiny{1/2  1   3/2}}}=1\,$,
$\,C^{\mbox{\tiny{1   1/2  1/2}}}_{\mbox{\tiny{1  -1/2  1/2}}}=\sqrt{2/3}\,$,
$\,C^{\mbox{\tiny{3/2  1   3/2}}}_{\mbox{\tiny{1/2   1   3/2}}}=-\sqrt{2/5}\,$,
$\,C^{\mbox{\tiny{1   1/2  3/2}}}_{\mbox{\tiny{1   -1/2  1/2}}}=1/\sqrt{3}\,$,
$\,C^{\mbox{\tiny{3/2  1  5/2}}}_{\mbox{\tiny{1/2 1  3/2}}}=\sqrt{3/5}\,$,
$\,R^{\mbox{\tiny{3/2}}}_{\mbox{\tiny{3/2,1/2,1/2}}}=-2/3\,$,
$\,R^{\mbox{\tiny{3/2}}}_{\mbox{\tiny{3/2,1/2,3/2}}}=\sqrt{5}/3\,$,
$\,R^{\mbox{\tiny{3/2}}}_{\mbox{\tiny{3/2,3/2,3/2}}}=2/3\,$,
$\,R^{\mbox{\tiny{3/2}}}_{\mbox{\tiny{3/2,3/2,1/2}}}=\sqrt{5}/3\,$, and
$\,R^{\mbox{\tiny{3/2}}}_{\mbox{\tiny{5/2,3/2,3/2}}}=1\,$.

In terms of $W_1$, $W_2$, and $W_3$, the constants $A_w$, $B_w$,
and $C_w$ are written as
\begin{eqnarray}
A_w &=& \sqrt{\frac{1}{54}}(W_1-W_2), \\
B_w &=& \sqrt{\frac{5}{54}}(W_1-W_2), \\
C_w &=& \frac{W_3}{\sqrt{5}},
\label{abc}
\end{eqnarray}
which implies that if $A_w=B_w$ only total isospin $5/2$ contributes to the decay.
This happens, in particular, for the initial state of
$\left|D\right>=\left|K^-\pi^+\pi^+\right>$. Therefore, the initial state should have be a
mixture of states. Indeed, the fittings we will show suggest $W_1\neq W_2$ and $W_3$
smaller than $W_1$ or $W_2$.
\begin{figure}[!htb]
\centering
\includegraphics[scale=0.5]{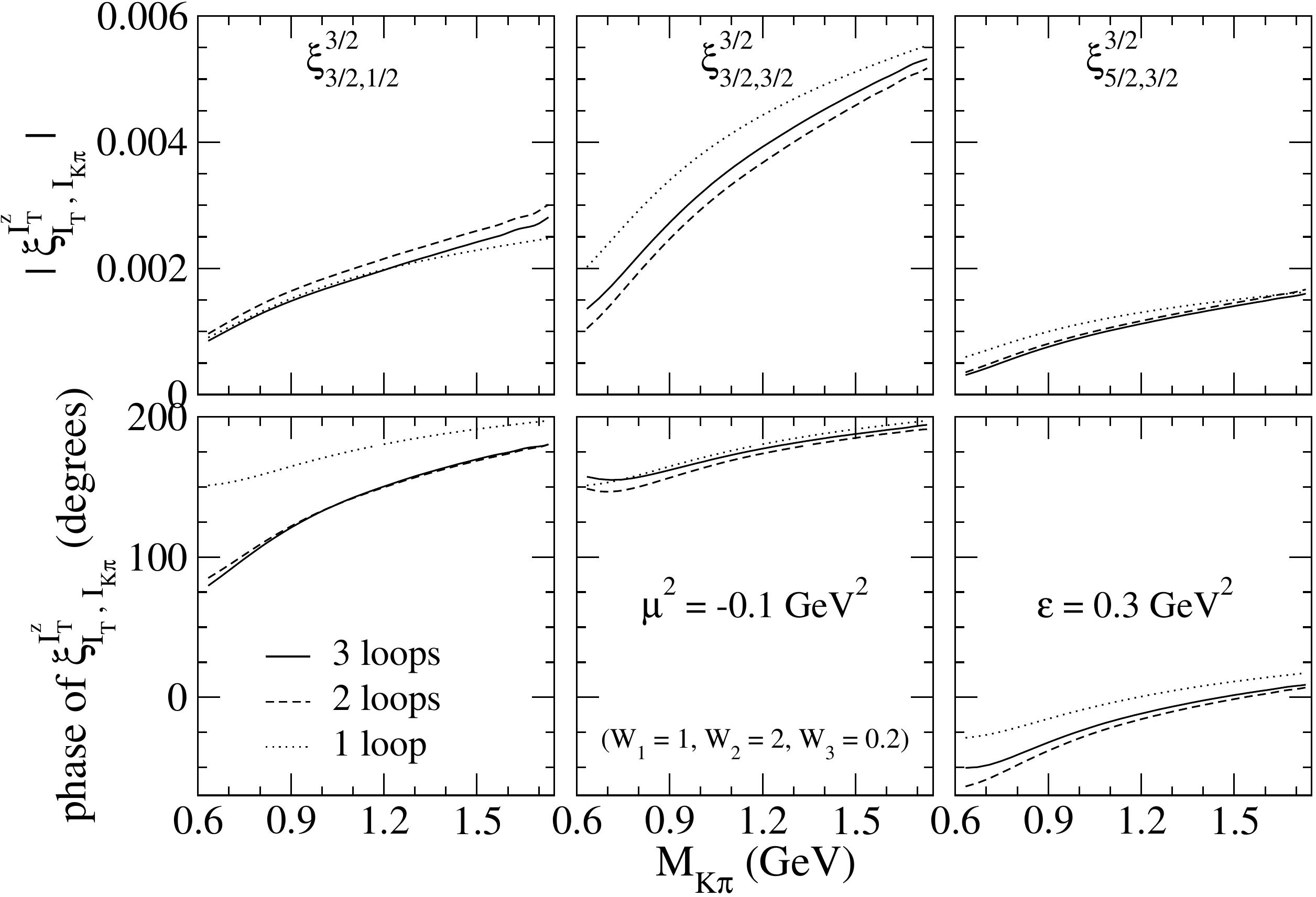}
\caption{Modulus and phase of $\xi^{I_T^z}_{I_T,I_{K\pi}}$
for $\varepsilon=0.3$ GeV$^2$ and $\mu^2=-0.1$ GeV$^2$. The parameters in the
the expansion of the initial state are $W_1=1$, $W_2=2$ and $W_3=0.2$.}
\label{qsicoupled}
\end{figure}

We compute up to three-loops the bachelor amplitude from the coupled equations for
$I_T=3/2$, Eq.~(\ref{eqint-coupled1}), and for the single channel equation for $I_T=5/2$,
Eq.~(\ref{eqint-coupled3}), with momentum cut-off of $2$~GeV. In Fig.~\ref{qsicoupled},
we show results for $\varepsilon=0.3$ GeV$^2$ and $\mu^2=-0.1$~GeV$^2$, with $W_1=1$,
$W_2=2$ and $W_3=0.2$. The convergence  $\xi^{I_T^z}_{I_T,I_{K\pi}}$ regarding the loop
expansion is evident, and two-loop calculations are enough for our purposes. The bachelor
amplitudes, present a considerable change in the phase and modulus, both increasing with
$M_{K\pi}$.

\section{Results for the Phase and Amplitude in the $D^+\to K^-\pi^+\pi^+$ decay}
\label{sec:results}

We will restrict our calculations up to two-loops as it was already
shown in Sec.~\ref{pertsol} to be enough to compute  bachelor amplitudes.
Results for two cases will be given, for the single channel model with interaction
restricted to $I_{K\pi}=1/2$, and the case where $I_{K\pi}=$ 1/2 and 3/2 interactions
are present in the $K\pi\pi$ system.

\subsection{Single-channel with $I_{K\pi}=1/2$ interaction}

\begin{figure}[!htb]
\centering
\includegraphics[scale=0.5]{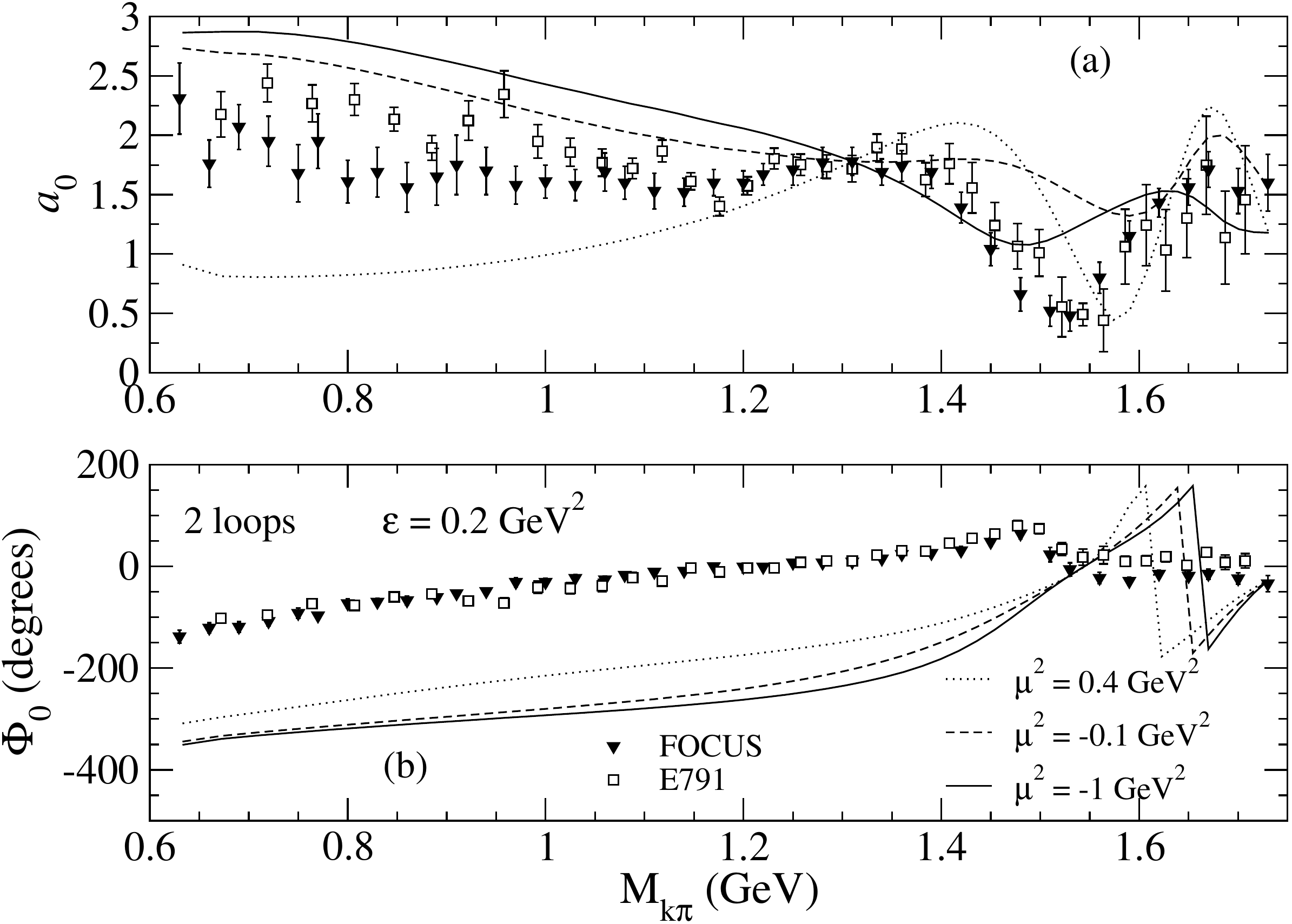}
\caption{(a) Modulus and (b) phase of the $D^+\to K^-\pi^+\pi^+$ S-wave amplitude
obtained from Eq.~(\ref{amplitude}). Values for $\mu^2$ in GeV$^2$:
$0.4$ (dotted line), $-0.1$ (dashed-line), and $-1$ (solid line). The data come from
the phase-shift analysis of E791 \cite{Aitala12,Aitala3} and FOCUS collaboration
\cite{FOCUS1,FOCUS2}.}
\label{amp-singlechannel}
\end{figure}

The physical amplitude for the s-wave $D^+\to K^-\pi^+\pi^+$  decay is obtained by
considering only $K\pi$ scattering in isospin $1/2$ states, with the bachelor amplitude
calculated by collecting the appropriate contributions up to two-loops
in Eq.~(\ref{itersol-single}). It is parametrized according to Eq.~(\ref{kpiamplf}) and
written as
\begin{small}
\begin{align}
A_0(M^2_{K\pi})
=\sqrt{\frac{2}{3}}\left[\frac{1}{12}\sqrt{\frac{2}{3}} +
\tau_{\mbox{\tiny{1/2}}}(M^2_{K\pi})\xi^{\mbox{\tiny{3/2}}}_
{\mbox{\tiny{3/2,1/2}}}(k_{\pi^\prime})\right].
\label{amplitude}
\end{align}
\end{small}
The modulus and phase of this amplitude is shown in Fig.~\ref{amp-singlechannel} and
compared to the experimental analysis from E791 \cite{Aitala12,Aitala3} and FOCUS
collaboration \cite{FOCUS1,FOCUS2}. The $K\pi$ isospin $1/2$ s-wave amplitude is fitted to
the LASS data in Sec.~\ref{sec:kpiampl}. To obtain the bachelor amplitude a small and
finite imaginary term ($\epsilon=0.2$ GeV) was introduced in the three-meson propagator,
it also represents absorption to other decay channels, which is beyond the model. An
arbitrary normalization point was chosen for Eq.~(\ref{amplitude}). Even though, there is
some sensitivity to the subtraction scale $\mu$ of the driving term, but as already
concluded in Ref.~\cite{MagPRD11}, no fittings to the data was found.

\begin{figure}[!htb]
\centering
\includegraphics[scale=0.5]{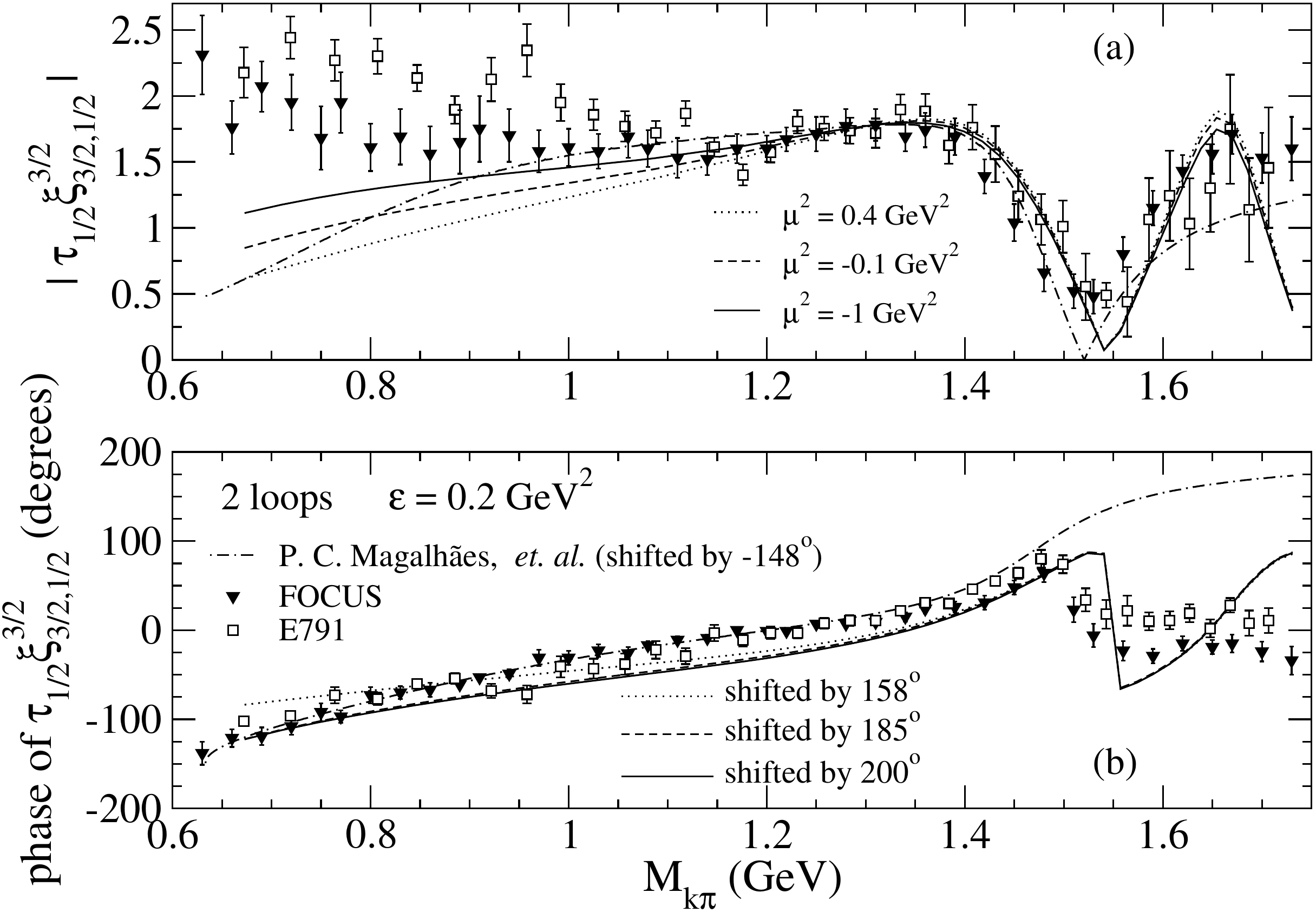}
\caption{(a) Modulus and (b) phase of
$\tau_{\mbox{\tiny{1/2}}}\xi^{\mbox{\tiny{3/2}}}_{\mbox{\tiny{3/2,1/2}}}$. Values for
$\mu^2$ in GeV$^2$: $0.4$ (dotted line), $-0.1$ (dashed-line), and $-1$ (solid line). For
reference, the dotted-dashed line gives the previous covariant calculation up to two-loops
from \mbox{P. C. Magalh\~aes, \it{et. al.}} of Ref.~\cite{MagPRD11}. The data come from
the phase-shift analysis of E791 \cite{Aitala12,Aitala3} and FOCUS
collaboration~\cite{FOCUS1,FOCUS2}.}
\label{tauqsi-sinlgechannel}
\end{figure}

The fit found in Ref.~\cite{MagPRD11} below $K^*_0(1430)$ suggested that the partonic
amplitude has little overlap with the $K^\mp\pi^\pm\pi^\pm$ final state channel, i. e.,
the first term in left-hand-side of Eq.~(\ref{amplitude}) should vanishes. Here, we also
show in Fig.~\ref{tauqsi-sinlgechannel}, results computed only by considering
$A_0(M^2_{K\pi})\approx \tau_{\mbox{\tiny{1/2}}}(M^2_{K\pi})\xi^{\mbox{\tiny{3/2}}}_
{\mbox{\tiny{3/2,1/2}}}(k_{\pi^\prime})$. As in the previous work~\cite{MagPRD11}, a
better fit to the experimental data below $K^*_0(1430)$ is found, compared to the results
showed in Fig.~\ref{amp-singlechannel}. However, note that a structure in the phase is
seen in the model which incorporates $K^*_0(1630)$ and $K^*_0(1950)$, as also verified in
the LASS data. A better fit of the LASS data above $K^*_0(1430)$ seems necessary to find a
better agreement with the valley in the modulus and the structure in the phase, as well.
The conclusion is somewhat independent on the subtraction point, at least for those small
values given in the figure.
 \begin{figure}[!htb]
\centering
\includegraphics[scale=0.5]{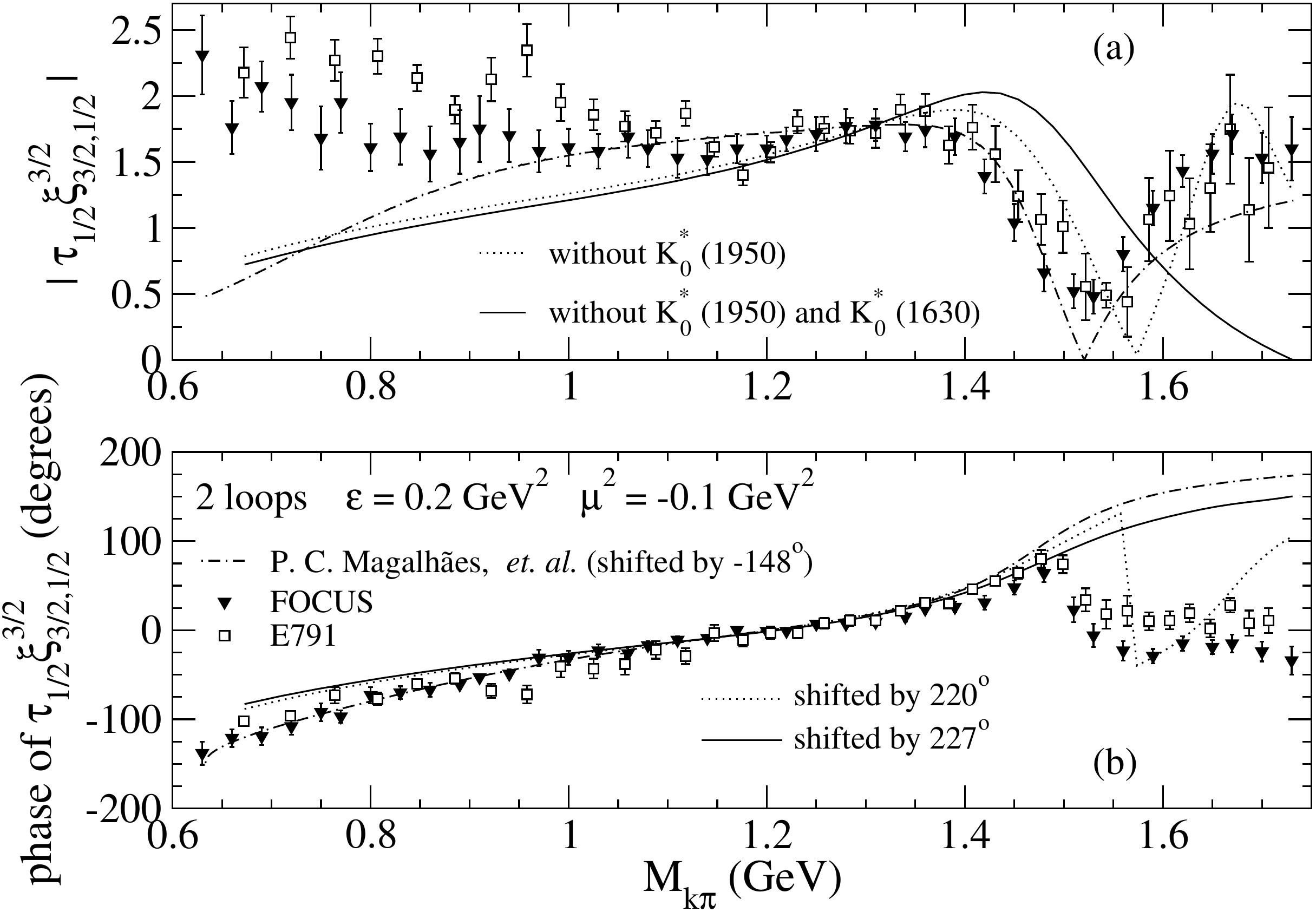}
\caption{(a) Modulus and (b) phase of
$\tau_{\mbox{\tiny{1/2}}}\xi^{\mbox{\tiny{3/2}}}_{\mbox{\tiny{3/2,1/2}}}$ for
two cases: i) without $K^*_0(1950)$ and ii) without $K^*_0(1950)$ and
$K^*_0(1630)$. Dotted-dashed line: theoretical calculation from \mbox{P. C. Magalh\~aes,
\it{et. al.}} of Ref.~\cite{MagPRD11}. Data from the phase-shift analysis of E791
\cite{Aitala12,Aitala3} and FOCUS collaboration~\cite{FOCUS1,FOCUS2}.}
\label{amp-sinlgechannel}
\end{figure}

In order to check the effect of the fitting to LASS data above $K^*_0(1430)$, we
remove from the $K\pi$ s-wave amplitude the $K^*_0(1950)$ and $K^*_0(1630)$ resonances,
as shown in Fig.~\ref{amp-sinlgechannel}. For this study, we fix the subtraction point at
$\mu^2=-0.1$ GeV$^2$. In the two sets of calculations, we turned off: (i) $K^*_0(1950)$
(dotted line), (ii) and both $K^*_0(1950)$ and $K^*_0(1630)$ (solid line). In case (i),
both the structure of the valley in the modulus and phase is somewhat kept, and make
distinct the results from Ref.~\cite{MagPRD11}, while in case (ii) as happens for the
reference calculation, the valley and mainly the phase, loose part of their structure. We
should note that for calculation (ii), the parameters of $\tau_{\mbox{\tiny{1/2}}}$ were
not refitted to the LASS data, and this can be observed by the shift in the valley
position of the modulus. Essentially, we restate that the on-shell
$K\pi$ amplitude should be represented well in order to compute the rescattering three-body
effects. Also, a simple fitting of the low-energy $K\pi$ amplitude without the detailed
physics of chiral symmetry, which leads to the broad $\kappa^*$ resonance, is somewhat
poor below $0.8$~GeV, as the figure suggests.

\subsection{Coupled-channels with $I_{K\pi}=$ 1/2 and 3/2  interactions}

We calculated the bachelor amplitudes iterating the coupled equations
(\ref{eqint-coupled1})-(\ref{eqint-coupled2}) and the single channel equation for total
isospin $5/2$, Eq.~(\ref{eqint-coupled3}), up to two-loops. In this case the amplitude for
the s-wave $D^+\to K^-\pi^+\pi^+$ decay is written as
\begin{eqnarray}
A_0(M^2_{K\pi})&=&C_1\left[\frac{A_w}{2}
+\tau_{\mbox{\tiny{1/2}}}(M^2_{K\pi})\xi^{\mbox{\tiny{3/2}}}
_{\mbox{\tiny{3/2,1/2}}}(k_{\pi^\prime})\right]
+C_2\left[\frac{B_w}{2}
+\tau_{\mbox{\tiny{3/2}}}(M^2_{K\pi})\xi^{\mbox{\tiny{3/2}}}
_{\mbox{\tiny{3/2,3/2}}}(k_{\pi^\prime})\right] + \nonumber \\
&+&C_3\left[\frac{C_w}{2}
+\tau_{\mbox{\tiny{3/2}}}(M^2_{K\pi})\xi^{\mbox{\tiny{3/2}}}
_{\mbox{\tiny{5/2,3/2}}}(k_{\pi^\prime})\right]
= a_0(M^2_{K\pi})e^{i\Phi_0(M^2_{K\pi})},
\label{amplitude-coupled}
\end{eqnarray}
where the constants $A_w$, $B_w$ and $C_w$ are defined in Eqs.~(\ref{awbwcw}).
The constants $C_i$ are given by
\begin{eqnarray}
C_1&=&\left<K^-\pi^+\pi^+\right|\left.I_T=3/2,I_{K\pi}=1/2,I_T^z=3/2\right>,\qquad \\
C_2&=&\left<K^-\pi^+\pi^+\right|\left.I_T=3/2,I_{K\pi}=3/2,I_T^z=3/2\right>,\qquad \\
C_3&=&\left<K^-\pi^+\pi^+\right|\left.I_T=5/2,I_{K\pi}=3/2,I_T^z=3/2\right>,\qquad
\label{kpiamplf2}
\end{eqnarray}
which comes from Eq.~(\ref{kpiamplf}). The driving terms of the integral equations for
$\xi^{I_T^z}_{I_T,I_{K\pi}}$, see Eqs.~(\ref{eqint-coupled1})-(\ref{eqint-coupled3}),
and the functional form of the amplitude given in Eq.~(\ref{amplitude-coupled}), depend
on only two free parameters, namely, $W_1-W_2$ and $W_3$. Actually, if we set $W_3=0$,
there are no free parameters anymore, since $W_1-W_2$ became an overall constant in the
amplitude.

\begin{figure}[!thb]
\centering
\includegraphics[scale=0.5]{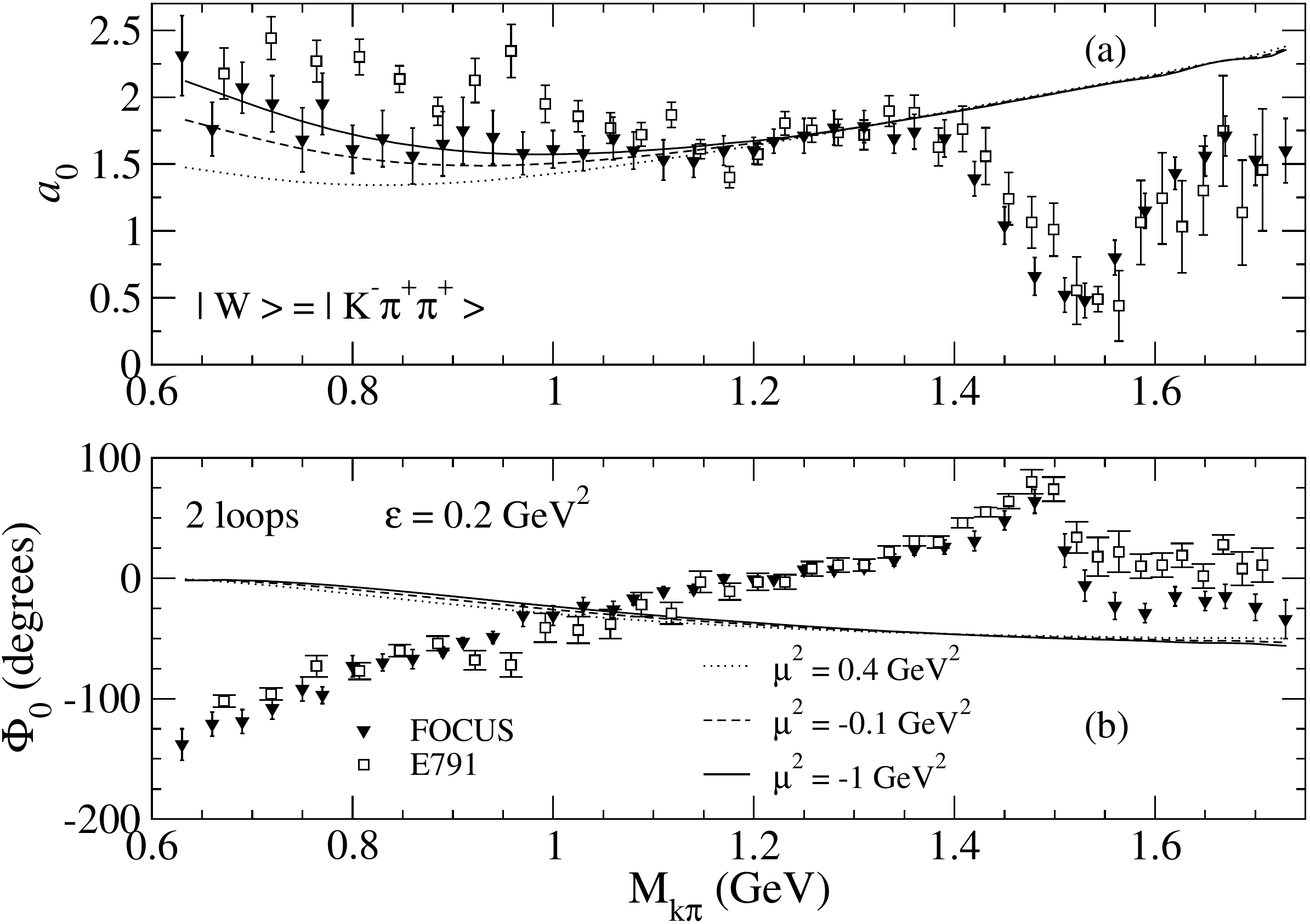}
\caption{(a) Modulus and (b) phase of the $D^+\to K^-\pi^+\pi^+$
amplitude for a  initial state with $W_1=W_2$ and $W_3=1$ in
Eqs.~(\ref{alpha1}). The data come from the phase-shift analysis
of E791 \cite{Aitala12,Aitala3} and FOCUS collaboration \cite{FOCUS1,FOCUS2}.}
\label{kpipi}
\end{figure}

The first striking result is that for $W_1=W_2$ and $W_3$ nonzero, which is also the case
for $\left|W\right>=\left|K^-\pi^+\pi^+\right>$ ($W_i=1$) is shown in Fig.~\ref{kpipi}.
Only total isospin $5/2$ is allowed and the $K\pi$ pair interacts in isospin $3/2$ state.
All the structure in the phase and amplitude is washed out, as the figure shows, excluding
that possibility as dominant for the partonic amplitude.
\vspace{0.3cm}

\begin{figure}[!thb]
\centering
\includegraphics[scale=0.5]{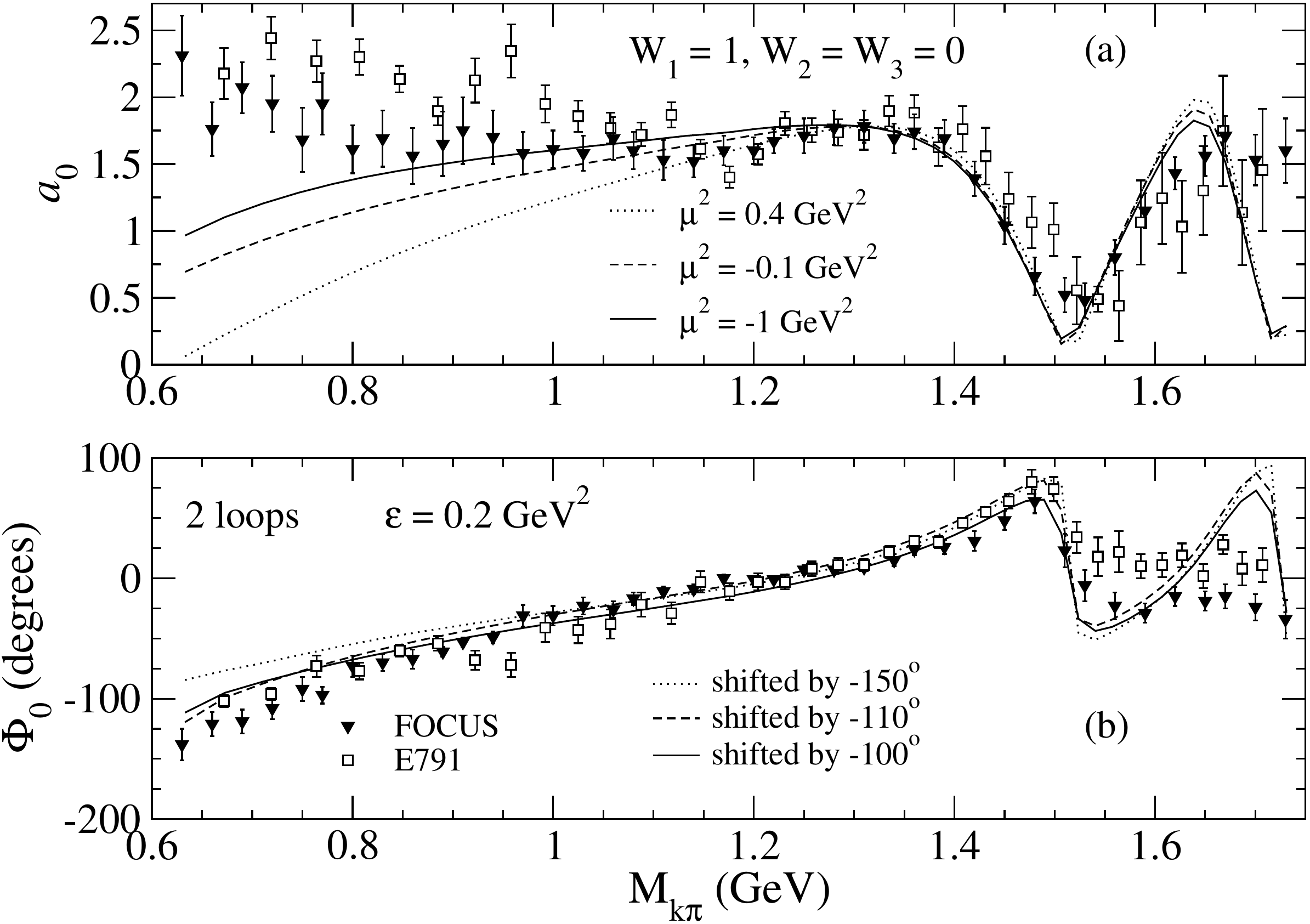}
\vspace{-0.5cm} \caption{Modulus (a) and phase (b) of the $D^+\to K^-\pi^+\pi^+$ amplitude for an
initial state with $W_1=1$, $W_2=W_3=0$ in Eqs.~(\ref{alpha1}). The data come from the
phase-shift analysis of E791 \cite{Aitala12,Aitala3} and FOCUS collaboration
\cite{FOCUS1,FOCUS2}.}
\label{w11w20w30}
\end{figure}

The relevant partonic weight $W_i$ should be guided by the difference $W_1-W_2$, which
means dominance of the total isospin $3/2$ in the initial state. In Fig.~\ref{w11w20w30},
we present results for $W_1=1$ and $W_2=W_3=0$, which corresponds
to a partonic amplitude given by
\begin{eqnarray}
\left|D\right>=\alpha^{\mbox{\tiny{3/2}}}_{\mbox{\tiny{3/2,1/2}}}\left|I_T=3/2,I_{K\pi}
=1/2,I^z_T=3/2\right> 
+\alpha^{\mbox{\tiny{3/2}}}_{\mbox{\tiny{3/2,1/2}}}\left|I_T=3/2,I_{K\pi^\prime}=1/2,
I^z_T=3/2\right> \ .
\end{eqnarray}
In the figure, we present results for $\mu^2=0.4,-0.1$ and $1$~GeV$^2$ and
$\epsilon=0.2$ GeV$^2$. A reasonable account of the experimental phase and modulus is
given by $\mu^2=-1$~GeV$^2$ and $\mu^2=-0.1$~GeV$^2$. At low $M_{K\pi}$ below 1 GeV,
the model does not describe the modulus, where the different analysis of E791 and FOCUS
present a large dispersion. The model tends to underestimate the modulus in the low mass
region. The bachelor amplitude increases with $M_{K\pi}$ (see e.g. Fig.~\ref{qsicoupled}),
which leads model to underestimate the modulus of the decay amplitude for low $K\pi$
masses. The characteristics valley and the follow-up height is somewhat described by the
model, with exception of the region close to the boundary of the decay phase-space, where
the data seems to indicates an increase of the amplitude and the model presents a
noticeable decrease.

\begin{figure}[!thb]
\centering
\includegraphics[scale=0.5]{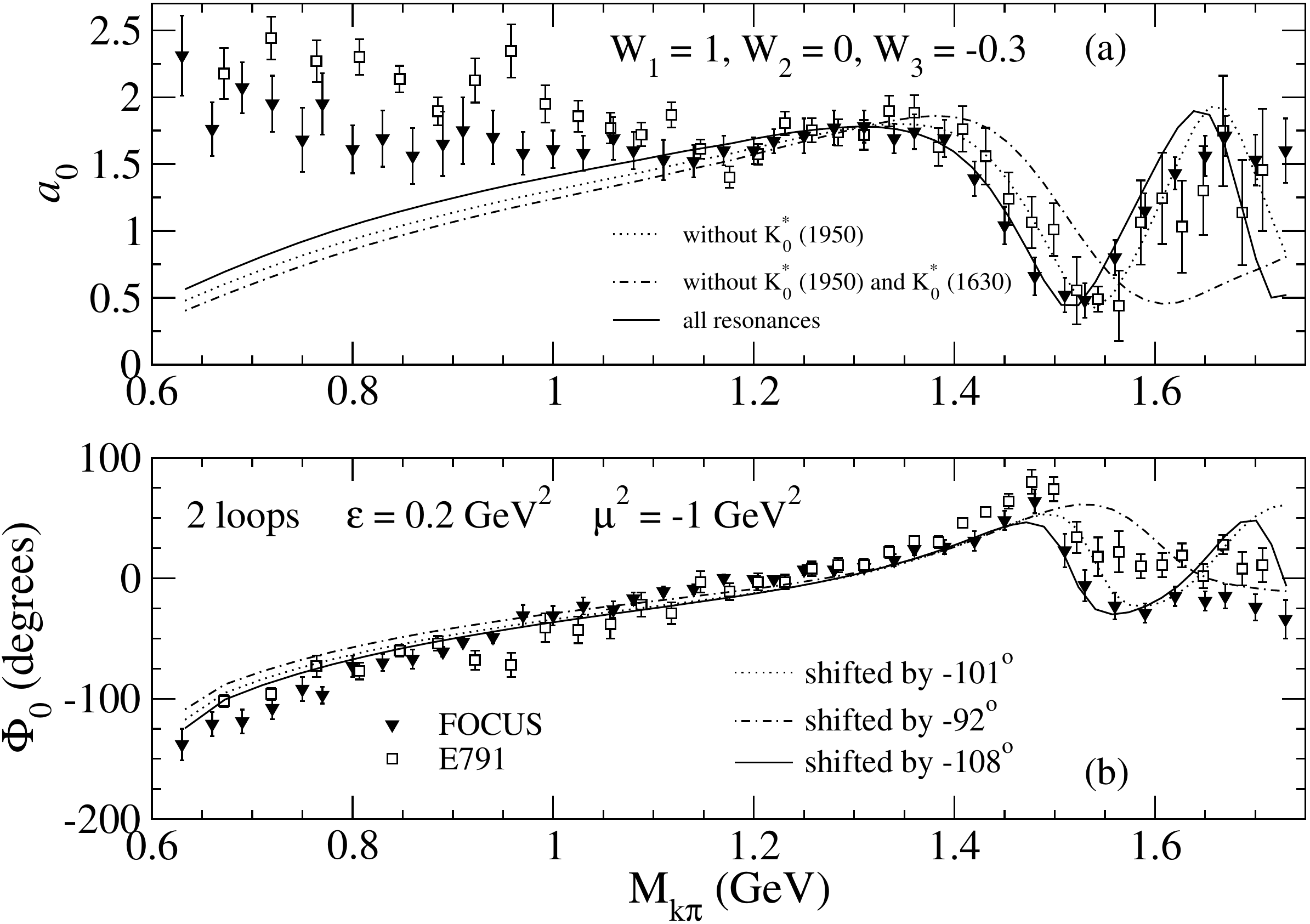}
\caption{(a) Modulus and (b) phase of the $D^+\to K^-\pi^+\pi^+$ amplitude for two cases: i)
without $K^*_0(1950)$ and ii) without $K^*_0(1950)$ and $K^*_0(1630)$. In all cases we
considered $W_1=1$, $W_2=0$ and $W_3=-0.3$. The data come from the phase-shift analysis
of E791 \cite{Aitala12,Aitala3} and FOCUS collaboration \cite{FOCUS1,FOCUS2}.}
\label{fittingsnores}
\end{figure}

We performed variations of the weight parameters and verified that a small mixture of
total isospin $5/2$ improves the fittings. We have used $W_1=1$, $W_2=0$ and
\mbox{$W_3=-0.3$} to obtain the results shown by the solid lines in
Fig.~\ref{fittingsnores} for $\mu^2=-1$~GeV$^2$. Notice also that the effect of the
resonances in the fit of the $K\pi$ isospin $1/2$ amplitude to the LASS data, in the last
model results, is similar to the single channel case we have already discussed. The
region close to the valley appearing in the modulus is sensitive mainly to our fit of the
LASS data in the neighborhood of $K^*_0(1630)$, while $K^*_0(1950)$ presents a smaller
effect in part due to the competition with the interaction in the $I_{K\pi}=3/2$ state.
The pronounced minimum  in the modulus of the decay amplitude, which appears in the $D-$decay
 phase-shift data at 1.53 GeV, should be contrasted with the LASS phase-shift in Fig. ~\ref{tau1232}, where
 the deep in not well pronounced and  placed at 1.65 GeV.

\section{Summary and Conclusions}
\label{conclusion}

We have investigated the three-body final state interaction effects in $D^+$ decays
focusing in the $K^-\pi^+\pi^+$ channel. In order to formulate the final state
interaction contribution to the decay, we used a relativistic three-body model for the
final state interaction in a heavy meson decay based on an approximation of the
Bethe-Salpeter-Faddeev equations  proposed in Ref.~\cite{MagPRD11} and generalized to
include different isospin channels of the interacting pair. The numerical calculations
were performed in three-dimensions, corresponding to the projection of the Bethe-Salpeter
like equations for the Faddeev components of decay amplitude to the light-front. We
generalized the quasi-potential approach applied to the light-front projection of the
Bethe-Salpeter equation to account for the three-body final state interaction in
heavy-meson decays. The calculations were performed with a truncated light-front equation
to the valence states and rotational symmetry was putted under control. The particular
kinematics of the decay in three-mesons, allows to choose the transverse plane as the
decay plane. This particular rotation around the z-direction is of kinematical nature and
therefore preserved by the truncation of the Fock-space.

The $K\pi$ S-wave amplitude model is fitted to the LASS data for isospin $1/2$, including
the resonances $K^*_0(1430)$, $K^*_0(1630)$ and $K^*_0(1950)$. The isospin $3/2$
amplitude is taken from an effective range formula already presented in Ref.~\cite{LASS}.
We allowed the partonic amplitude to have nonzero weight in the three possible isospin
states with $I_T$ equal to 3/2 and 5/2. A small contribution  of $I_T=5/2$ seems to
improve the fit of the data of amplitude and phase from E791 \cite{Aitala12,Aitala3} and
FOCUS \cite{FOCUS1,FOCUS2} collaborations.

We showed that the loop-expansion to calculate three-body rescattering effects in the
$K\pi\pi$ channel converges fast, and the solution of the integral equations for the
bachelor amplitudes by iteration at the three-loop level  gives a contribution that can be
neglected in respect to the two-loop results. We explored the dependence on the model
parameters in respect to the partonic amplitude.

We found that the negative value of the phase seen in the data
\cite{Aitala12,Aitala3,FOCUS1,FOCUS2}, can be obtained by an appropriate choice of the
real weights of the three isospin components of the partonic amplitude, with a small
mixture of total isospin 5/2. The feature of the modulus of the unsymmetrized decay
amplitude presenting a deep valley and a following increase, for $K\pi$ masses above 1.5
GeV, is fairly reproduced, which indicates an assignment of  $0^+$ to the isospin 1/2
$K^*(1630)$ \cite{PDG} omitted from the PDG summary table.   Below 1 GeV the
model underestimate the data for the
modulus, as happens close to the end of the available phase-space around 1.8 GeV.

Certainly, a better comprehension of the $K\pi$ amplitude in the physical and unphysical region, and in
particular above $K^*(1430)$ can  bring more realism to the description of the three-body final state interaction in
$D$ decays. The challenge of applying the    formalism to $B$ decays and CP
violation~\cite{LHCb1} by  extending Ref. \cite{CPT} to include three-body FSI,  is let to a future work.

\acknowledgments
We thank the Brazilian funding agencies FAPESP
(Funda\c{c}\~{a}o de Amparo a Pesquisa do Estado de S\~{a}o Paulo) and
  CNPq (Conselho Nacional de Pesquisa e
Desenvolvimento of Brazil). We are grateful to I. Bediaga, P.  C. Magalh\~aes and M. Robilotta for the discussions.


\end{document}